\documentclass[referee]{raa}            

\usepackage{graphicx,times}             
\usepackage{natbib}
\usepackage{booktabs}
\usepackage{amssymb,amsmath}
\usepackage{subcaption}
\usepackage{multirow}

\bibpunct{(}{)}{;}{a}{}{,}

\usepackage[pagebackref=true]{hyperref}

\begin{document}

\title{The GECAM Real-Time Burst Alert System}

   \volnopage{Vol.0 (20xx) No.0, 000--000}      
   \setcounter{page}{1}          

   \author{Yue Huang* 
      \inst{1}
    \and Dongli Shi
      \inst{1,2}
    \and Xiaolu Zhang
      \inst{1,3}
   \and Xiang Ma
      \inst{1}
    \and Peng Zhang
      \inst{1,2}
    \and Shijie Zheng
      \inst{1}
   \and Liming Song
      \inst{1}
    \and Xiaoyun Zhao
      \inst{1}
    \and Wei Chen
      \inst{1}
    \and Rui Qiao
      \inst{1}
    \and Xinying Song
      \inst{1}
    \and Jin Wang
      \inst{1}
    \and Ce Cai
      \inst{1,4}
    \and Shuo Xiao
      \inst{1,4}
    \and Yanqiu Zhang
      \inst{1,4}
    \and Shaolin Xiong*
      \inst{1,4}
   }
   \institute{Key Laboratory of Particle Astrophysics, Institute of High Energy Physics, Chinese Academy of Sciences, Beijing 100049, China; {\it huangyue@ihep.ac.cn, xiongsl@ihep.ac.cn}\\
        \and
             Southwest Jiaotong University, Chengdu 611756, China\\
        \and
             Qufu Normal University, Qufu 273165, China\\
        \and
            University of Chinese Academy of Sciences, Chinese Academy of Sciences, Beijing 100049, China\\
\vs\no
   {\small Received 20xx month day; accepted 20xx month day}}

\abstract
{Gravitational Wave High-energy Electromagnetic Counterpart All-sky Monitor (GECAM), consisting of two micro-satellites, is designed to detect gamma-ray bursts associated with gravitational-wave events. Here, we introduce the real-time burst alert system of GECAM, with the adoption of the BeiDou-3 short message communication service. We present the post-trigger operations, the detailed ground-based analysis, and the performance of the system. In the first year of the in-flight operation, GECAM was triggered by 42 GRBs. GECAM real-time burst alert system has the ability to distribute the alert within $\sim$1 minute after being triggered, which enables timely follow-up observations.
}
   \authorrunning{Yue Huang, Dongli Shi, Xiaolu Zhang, et al }            
   \titlerunning{GECAM Real-Time Burst Alert System}  

   \maketitle

\section{Introduction} \label{sec:intro}

On September 14, 2015, the first detection of gravitational wave (GW) signals from the merger of two stellar-mass black holes, observed by the Laser Interferometer Gravitational-Wave Observatory (LIGO) detectors, inaugurated the era of GW astronomy \citep{2016PhRvL.116f1102A}. This was the first direct evidence of the predictions of general relativity. On August 17, 2017, the Advanced LIGO and Advanced Virgo Gravitational-Wave interferometers detected the first GW, GW 170817, from a binary neutron star merger, significantly promoting the study of gravitational-wave multi-messenger astronomy \citep{2017PhRvL.119p1101A}. \emph{Fermi} and \emph{INTEGRAL} detected a short gamma-ray burst (GRB), GRB 170817A, 1.7 s after the GW events. The electromagnetic (EM) follow-up observations not only succeeded in localizing the merger to the host galaxy, NGC 4993, but also provided the first unambiguous detection of a kilonova, the broadband signature of rapid neutron capture nucleosynthesis (r-process) in the merger ejecta. These detections made by GW and EM observatories, for the first time, validated the merger model proposed decades ago to explain the short GRBs \citep{1986ApJ...308L..43P}.

The identification of EM counterparts to GW events allows for the precise localization of the GW source, which would further yield rich scientific rewards (see \citealt{2020PhR...886....1N} for a review). The EM counterpart identification is constrained by the accuracy of the localization of the GW signal, which is usually expected to be a few hundreds of square degrees \citep{2020LRR....23....3A}. In general, we expect that searching for high energy EM counterparts to a GW event will play a major role in the discovery of the EM counterpart. This is because, firstly, the luminosity of the high-energy counterpart is large and less likely to be absorbed by the medium; secondly, in the low energy bands, there might be few optical candidates localized within the error region of the GW source (i.e., \citealt{2016ApJ...826L..13A}). Since the high energy sky is less "crowded", it is more reasonable to relate a high energy transient to the GW event; thirdly, the time delay between the high-energy emission and the GW emission is assumed to be minimal. Therefore, a precise localization of the high-energy transient could substantially reduce the localization uncertainty of the GW event, which further facilitates the follow-up observations at other wavelengths. In recent years, a large number of observations have been made with hard X-ray and $\gamma$-ray telescopes, such as \emph{Fermi}-GBM \citep{2009ApJ...702..791M}, \emph{Swift}-BAT \citep{2005SSRv..120..143B}, \emph{INTEGRAL}-SPI-ACS \citep{2003A&A...411L...1W}, \emph{Insight}-HXMT \citep{2020SCPMA..6349502Z,2021MNRAS.508.3910C} and Konus-\emph{Wind} \citep{1995SSRv...71..265A}, to search for high energy counterparts to GW sources.

Gravitational wave high-energy Electromagnetic Counterpart All-sky Monitor \citep{2020SSPMA..50l9508L,2021arXiv211204772L} (GECAM, also known as ``HuaiRou-1'') is a space-based project proposed for the detection of high-energy EM counterparts to GW sources, as well as other high-energy transient sources, i.e., GRBs and magnetars. GECAM consists of two micro-satellites, GECAM-A and GECAM-B, which are designed to operate on identical orbits (600 km altitude and 29$^{\circ}$ inclination), on opposite sides of the Earth, in order to get a simultaneous view of the entire sky. Each satellite features a dome-shaped array of 25 Gamma-ray detectors (GRD) and 8 Charged particle detectors (CPD). The GRDs are composed of a LaBr$_3$ crystal and silicon photomultiplier tube (SiPM) array, covering an energy range from 6 keV to 5 MeV \citep{2021arXiv211204774A}. The CPDs are used to monitor the flux of charged particles on the GECAM orbit and help distinguish between astrophysical events and charged particle events. The CPDs use plastic scintillators combined with SiPM, covering an energy range of 300 keV--5 MeV \citep{2021arXiv211205314X}. In case of a trigger, the flight software \citep{2021arXiv211205101Z} catches the in-direction and provides a preliminary classification to the source, which will be downlinked as a trigger alert to the ground. In order to carry out rapid follow-up observations at other wavelengths, a real-time downlink of the alert data is required. Considering the current status of the real-time downlink resources in China, GECAM adopts the global short message communication service \citep{2021AdSpR..67.1701L} of BeiDou-3 navigation satellite system \citep{2019Navigation..66.7Y} to downlink the trigger alert data to the ground. GECAM is the first satellite to use the BeiDou-3 global short message service on board and the first space astronomy satellite in China capable of real-time downlink.

The GECAM Scientific Ground Segment \citep{2020SSPMA..50l9512C,2021RAA_inprep_Z} thus includes a section that is devoted to process the BeiDou short messages upon their arrival. In the following, we describe the onboard triggering and data flow in Section \ref{sec:RTS} and the real-time burst alert system in Section \ref{sec:DA}. We report the in-flight performance of the first year in Section \ref{sec:IOP}. At last, in Section \ref{sec:CS} we give a summary.

\section{Onboard Triggering and Data Flow} \label{sec:RTS}

\subsection{In-flight Trigger and Localization} \label{sec:IFTL}

The GECAM In-flight Realtime Trigger and Localization software (GIRTLS) \citep{2021arXiv211205101Z} continuously monitors the background count rates of all GRDs for significant increases on different energy ranges and timescales, to detect GRBs and other short-timescale transients. The background is accumulated over 20 s pre-trigger, excluding the most recent 5 seconds (in default). The event data are binned to 50 ms and 8 energy channels, which means that the trigger timescales are defined as multiples of 50 ms until reaching 4 s. Except for the 50 ms timescale, all of the triggers include two phases offset by half of the time bin. GECAM supports 64 different trigger algorithms, each of which comes with an adjustable threshold. The trigger algorithms currently implemented include five energy ranges and seven timescales, a detailed description of the 64 algorithms can be found in \cite{2021arXiv211205101Z}. A trigger is only generated when at least three detectors exceed the threshold at the same time. When there is a trigger, the GIRTLS gives an approximate location to the source using the relative rates recorded in the 25 GRDs that accumulated on 4 timescales. Besides GRBs, there are other events, such as solar flares and charged particle events that can trigger the alert, so the GIRTLS further performs a classification by using the count ratio between CPD and GRD, the localization, the hardness ratio, and the geographic location of the satellite to identify the type of source.

Once triggered on board, the GIRTLS produces the trigger alert data that are downlinked to the ground via BeiDou short message. The trigger alert data includes information on the trigger significance, the burst spectrum, on-board localization and classification, and light curve for improving ground localization. There are two algorithms to localize the burst on the ground: one using the relative count rates from the 25 GRDs, which requires a relatively long-time light curve from each detector; the other one using the time delay of the burst between the two satellite, which operates on high temporal resolution light curves \citep{2021ApJ...920...43X}. Due to the limitation of capacity of single BeiDou short message (560 bits per message) and downlink capacity of the BeiDou system \citep{2021AdSpR..67.1701L}, the high temporal resolution light curve is only generated for short bursts that are believed to be related to neutron star mergers \citep{1986ApJ...308L..47G}.

There are two types of trigger alert data: long trigger and short trigger. If the count rate exceeds the threshold at 4 s and 20 s post-trigger, the trigger will be identified as long trigger. Each long trigger is comprised of 31 BeiDou short messages. The first two messages contain the most important parameters for the rapid follow up observations, i.e., trigger time, burst localization, classification and spectrum, satellite position and attitude at trigger time, with backups. The 3rd and 4th messages contain light curves from three GRDs with the highest and lowest trigger significance, which is binned by different trigger timescales and energy ranges. The light curves provide a quick view of the burst. The 5th message contains the light curve of 8 CPDs, covering from 30 s prior to 180 s following the trigger time, which is used to distinguish particle events from GRBs. The 6th to 30th messages store the light curves from each GRD from $\sim$50 s before the trigger (divided into 8 time bins) to 185 s after the trigger (divided into 22 time bins) and are binned by timescales from 50 ms to 50 s, with shorter timescales close to the trigger time. The last message gives the satellite attitude which lasts 120 s after the trigger time. The BeiDou short messages transmit every 17 s and take about 10 minutes to finish all 31 messages.

The difference between the short and long trigger alert data is that the short trigger includes a combined high-resolution (0.4 ms in default) light curve from 25 GRDs with 2500 bins. Each short trigger contains up to 31 short messages, depending on the size of the light curve after compression. The first two messages are the same as the long trigger. The rest of the messages are the compression method and the compressed light curve.

\subsection{On-ground Analysis} \label{sec:OGA}

After being received by the National Space Science Center (NSSC) on ground, the BeiDou short message is forwarded to Scientific Ground Segment at Institute of High Energy Physics (IHEP) and ingested into the Burst Alert System (BAS). The BAS is developed to process the trigger alert data in real-time and transmit the locations and other important information to the astronomy community via the standard communication channel (e.g., the GRB Coordinates Network (GCN) \footnote{See http://gcn.gsfc.nasa.gov/gcn}). The types of GECAM notices generated by the BAS are listed below.

1. \textbf{GECAM FLIGHT}: trigger time, trigger energy range, trigger significance, on-board localization (RA and Dec), ground refined classification (see Section \ref{sec:31}), $\sim$1 minute after trigger.

2. \textbf{GECAM GROUND}: ground localization (RA and Dec, see Section \ref{sec:32}) and classification (see Section \ref{sec:31}), $\sim$10 minutes after trigger.

The notices are sent only if the BAS classified the trigger as an astrophysical transient, such as a GRB. Since July 15, 2021, we sent a total of 323 notices in 2021, of which 156 were flight and 167 were ground, containing 205 triggers.

The BAS provides a refined classification by using an updated algorithm (see Section \ref{sec:31}). Due to the limitation on memory and computational resources on board, the GIRTLS uses a coarser sky grid (3072 grid points), three pre-defined templates (soft, normal and hard spectra in Band function), and an averaged pre-burst background level to localize the source. Compared to GIRTLS, the BAS provides improved locations by applying a finer sky grid, fitting the burst spectrum, and estimating the background with pre- and post-trigger data (see Section \ref{sec:32}) or with the time delay calculated based on the Modified Cross-correlation Function \citep[Li--CCF, ][]{2021ApJ...920...43X} when a burst is observed by both satellites, or GECAM and other satellites (see Section \ref{sec:33}).

Moreover, GECAM produces time-tagged event data that are transmitted via the X-band ground station. The X-band data are not downlinked in real-time like the alert data, but delayed up to several hours based on the passages over the station. The X-band data are used to determine the final characteristics of the bursts. The continuous event data also enhances the ground-based searching for untriggered GRBs by using the coherent search method, which was initially applied to \emph{Insight}-HXMT \citep{2021MNRAS.508.3910C}.

\section{The Burst Alert System (BAS)} \label{sec:DA}

\subsection{Re-classification of the trigger} \label{sec:31}

GECAM will detect GRBs, solar flares, particle events, soft gamma repeaters (SGRs) and earth occultation of bright sources (e.g., Sco X--1). The GIRTLS in-flight uses the background-subtracted counts ratio between CPD and GRD to identify particle events and further uses the event localization (the error box is 2 $\sigma$) and hardness ratio to distinguish known sources. Hence, it is only valid when the background is correctly estimated and a precise location is obtained.

On the other hand, the BAS on-ground provides a refined classification to each trigger. The relevant data applied are event localization, hardness ratio, count rate of CPD, count ratio of CPD and GRD, the location of the spacecraft, and McIlwain magnetic L coordinates. Particle events occur predominantly in trapped particle regions, mostly in the entry or exit of the South Atlantic Anomaly (SAA) region, or at high L values. Thus, they are identified when three of the following four conditions are met: spacecraft geographic location, L value, CPD count rate and the count ratio between CPD and GRD. Like in GIRTLS, the BAS compares the event location with the sun and other known sources, e.g. SGR 1935+2154, with the error box set to 3 $\sigma$ of the location error and includes the systematic error. If the hardness ratio is in the predefined range, and the source (the sun and other known sources) is not occulted by Earth, the event is classified as a solar flare or burst from known sources. Events which are located near the galactic plane and have a hardness ratio above one will be classified as generic sources. GECAM can also be triggered by bright sources rising from the Earth's limb, and this can be easily identified since the occultation time for each source can be calculated precisely.

\subsection{Ground localization using relative rates} \label{sec:32}

\subsubsection{Background estimation} \label{sec:321}

\begin{figure*}
\centering
    \includegraphics[scale=0.7]{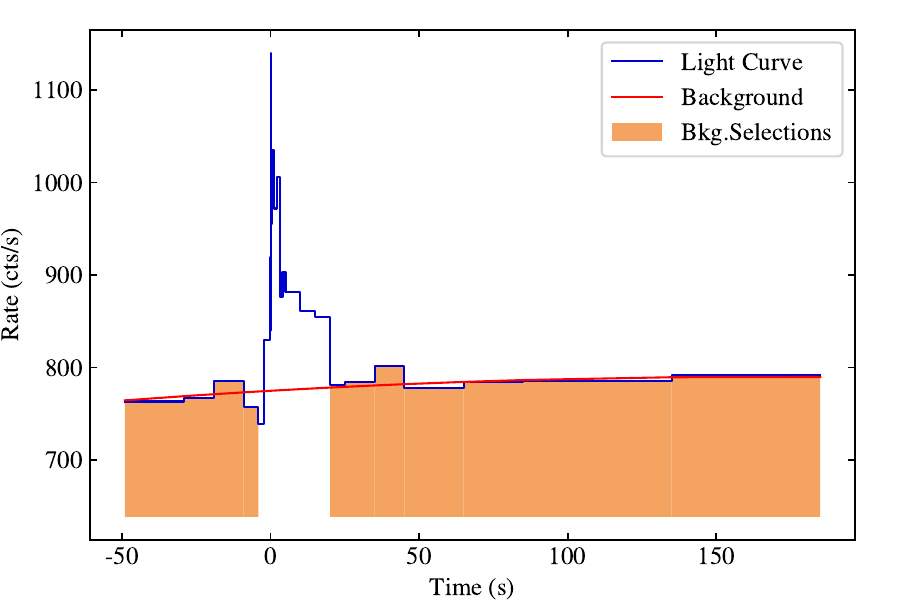}
    \hspace{1in}
    \includegraphics[scale=0.75]{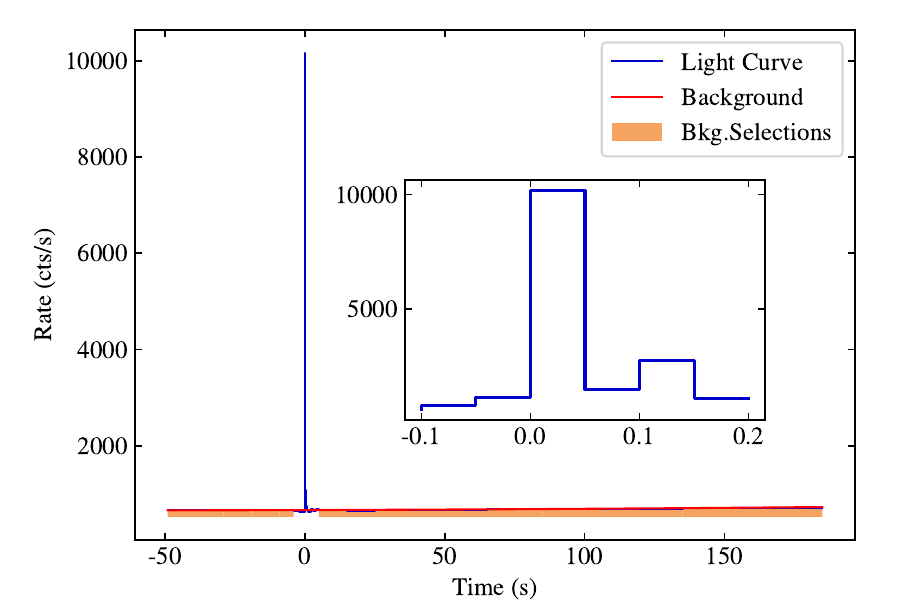}
    \caption{Background selection for GRB 211102B (upper panel) and a bright burst from SGR J1935+2154 (T0=2021-09-11 05:32:38.65 UT, lower panel). Shown is the data of one GRD in the 15--1020 keV energy range. The orange filled region shows the data chosen to perform the fit and the red line is the background estimate. }
    \label{fig:bkg}
\end{figure*}

The BAS performs background fittings after the BeiDou short messages are complete. The method applied here is recursive non-parametric regression, similar to what is adopted by \emph{Fermi}-GBM RoboBA \citep{2020ApJ...895...40G}. First, we fit the data from -49.1 to -4.1 s (divided into 4 time bins, binned by timescales from 5 to 20 s) pre-trigger and 5 to 185 s (divided into 10 time bins, binned by timescales from 5 to 50 s) post-trigger by a polynomial function up to second order for each GRD, respectively. When at least four detectors exceed the predefined signal-to-noise ratio thresholds, the corresponding bins will be removed from the background. The regression will perform repeatedly on the remaining time bins, until the recursive process converges (see Figure \ref{fig:bkg}). When there are less than two bins at pre-trigger or post-trigger, the BAS cannot perform background fitting, and the background is thereby averaged by pre-trigger. This usually happens during extreme background fluctuation, i.e., the satellite is close to SAA, or when the burst duration is abnormally long. There are 6 out of 37 GRBs \footnote{GECAM detected 42 GRBs in 2021, but 5 of which dropped the data packets. see Section \ref{sec:41}} which failed to fit the background. Five failures result from the long burst duration, the other one is caused by the background fluctuation. The BAS has a success rate of about 84\%.

\subsubsection{Spectrum fit and localization} \label{sec:321}

\begin{figure}
\centering
\includegraphics[scale=0.7]{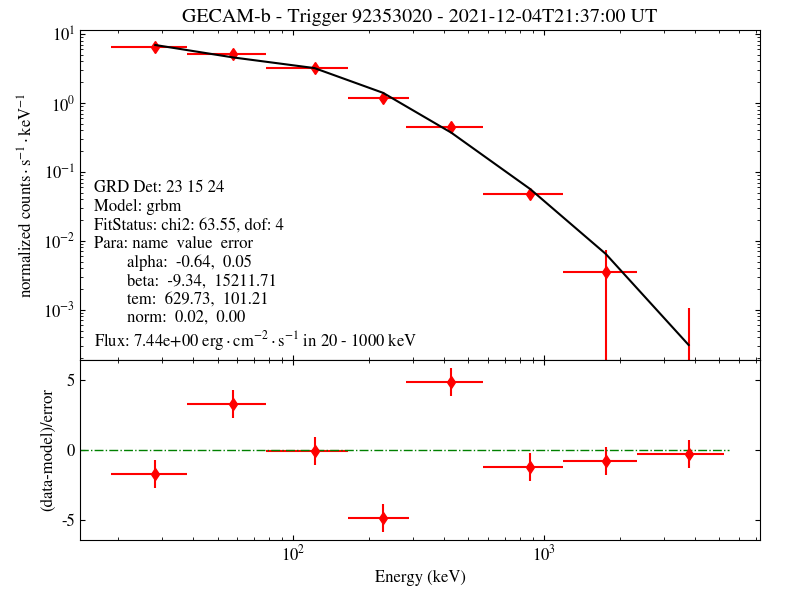}
\caption{An example (GRB 211204C) of the burst spectrum. Data from GRD 15, 23, and 24 are used. The best-fit model (black solid line) and data are shown in the upper panel, the residuals are shown in the lower panel. }
\label{fig:spec_ex}
\end{figure}

\begin{figure*}
\centering
\includegraphics[scale=0.5]{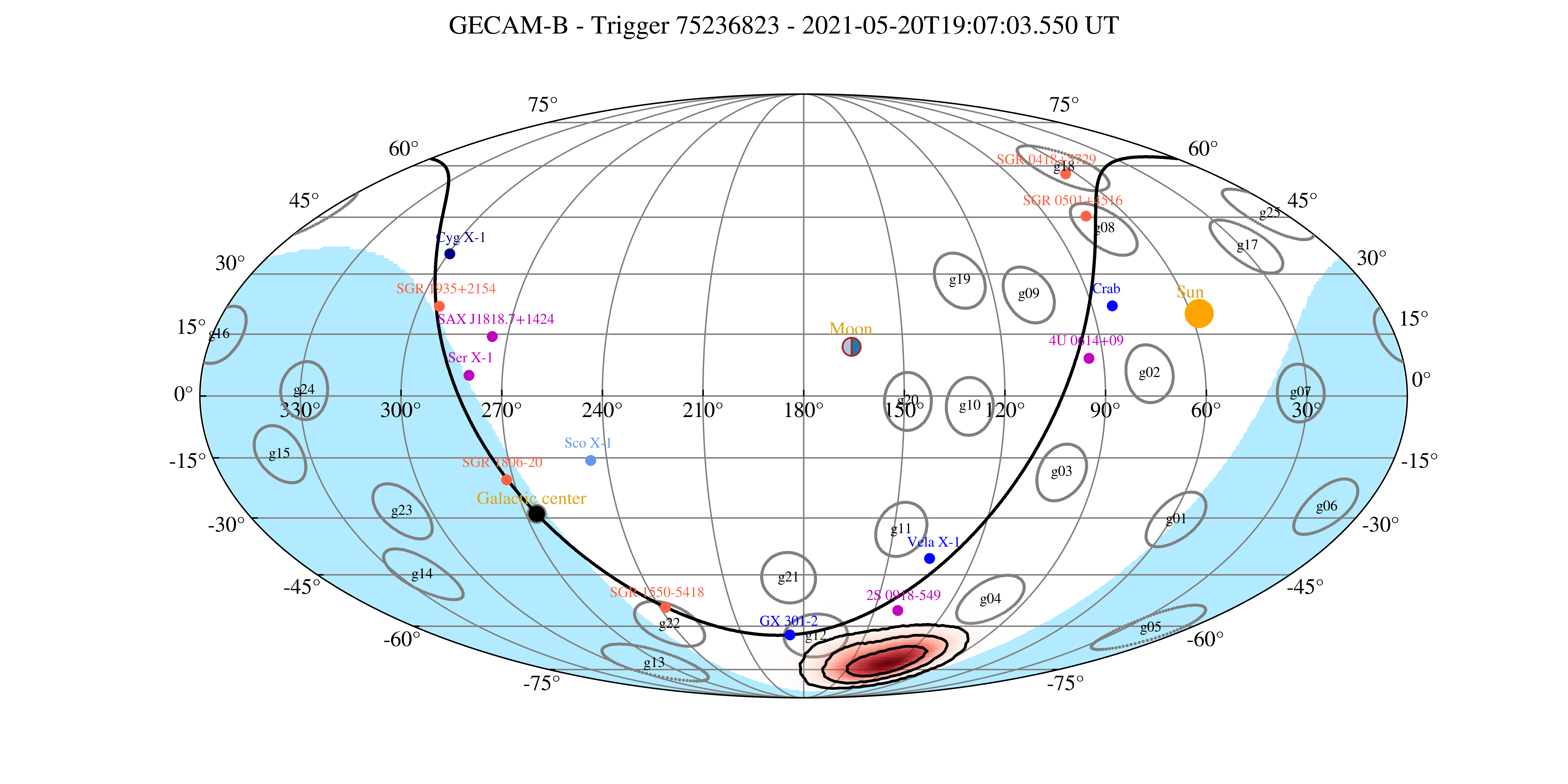}
\caption{An example (GRB 210520A) of the full sky map of the localization. The localization posterior is shown with a red gradient. The detector pointings at the time of the trigger are shown as light gray circles for all 25 GRDs (Note that the size of the circles do not represent the field of view of the detectors, only the pointing of the detector normal). The Galactic plane is shown as a black line with a circle denoting the Galactic center. The Earth occultation is shown in blue, and the Sun and moon are shown in yellow and blue-gray, respectively. Additionally, several bright sources are shown in different colors.}
\label{fig:loc_ex}
\end{figure*}

The GECAM on-board localization system operates on three spectral templates, which leads to an inaccurate localization if the spectral templates mismatch with the actual spectrum. Ideally, this can be corrected by simultaneously fitting the spectrum and location \citep{2018MNRAS.476.1427B}. However, the small number of time and energy bins of the trigger alert data are not suitable for this fitting. Thus, one needs to fit the spectrum and location iteratively. The burst spectrum is a combination of the 3 detectors with the highest trigger significance. These detectors usually have a similar incidence angle and therefore response. We added their response files. First, we generate the response file using the on-board location and fit the spectrum with the Band function and cut-off power-law model (see Figure \ref{fig:spec_ex}). Then, we construct the template for each detector in 15--1020 keV range over 12,288 grid points in the payload coordinates, with the best-fitting model and parameters. These are compared to the observed counts accumulated in the 25 GRDs, to find a $\chi ^2$ minimum. And the position is converted to equatorial coordinates using the spacecraft attitude. The new position is used as input for the next iteration, until the position converges. A full sky HEALPix map of the localization is then produced, see Figure \ref{fig:loc_ex} for an example.


\subsection{Ground localization using Time delay method} \label{sec:33}

In addition to the spectral fitting method, GRBs can also be located via the time delay method or triangulation technique \citep{1997ApJS..110..157L}. When a GRB arrives at two spacecrafts, it can be localized to an annulus characterized by the time delay and spacecraft positions. The time delay and its uncertainty are usually calculated by the cross-correlation function (CCF). However, when the classic CCF method is applied to locate GRBs for low orbit satellites, the localization region becomes too large to give effective constraints. To make an improvement, \cite{2021ApJ...920...43X} proposed an improved time delay localization method based on a Modified Cross-correlation Function (MCCF, Li--CCF) \citep{1999ApJ...521..789L}, from which it provides an accurate time delay from the high time resolution light curves.

\begin{figure}
\centering
\includegraphics[scale=0.7]{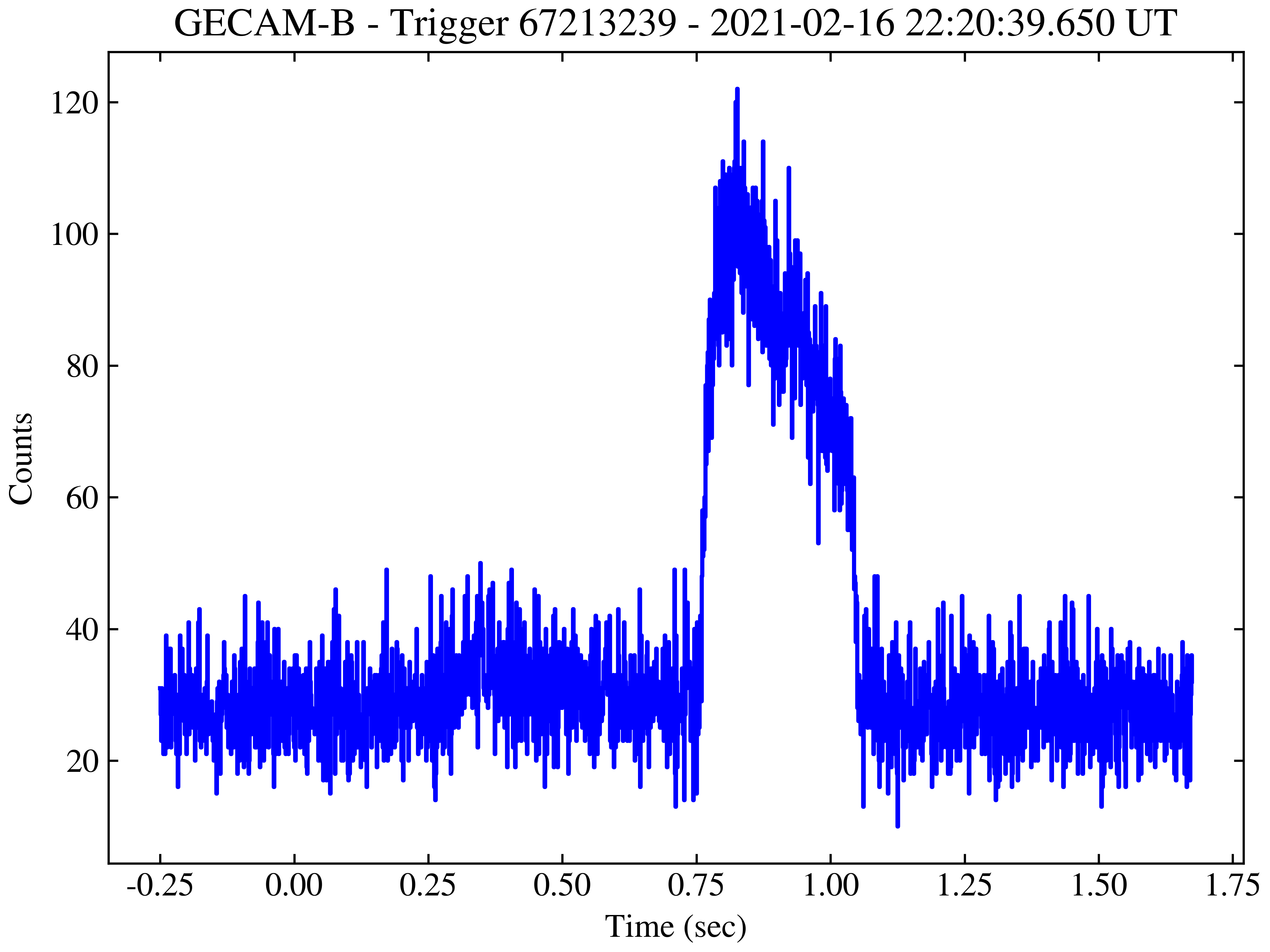}
\caption{An example (SGR J1935+2154, T0=2021-02-16 22:20:39.650 UT) of the high time resolution (1 ms) light curve. Shown is the sum of the data from all GRDs.}
\label{fig:hrlc}
\end{figure}

Once all the short trigger alert data are received, BAS decompresses it to obtain a high time resolution light curve (see Figure \ref{fig:hrlc}). If a burst is observed by both satellites, the light curves are sent to the MCCF localization algorithm. \cite{2021ApJ...920...43X} provides a full description of the algorithm and an estimate of the uncertainty (1$\sigma$: less than 0.3 $^{\circ}$). Consequently, the annulus is excluded by the Earth occulted part and combined with the localization derived by comparing the count rates from different detectors \citep{2022MNRAS.tmp..994X}.

Because GECAM-A has not turned on yet (see Section \ref{sec:IOP}), there are no GRBs or other bursts that have been localized with this method by the two GECAM satellites. However, we have applied this method to locate a burst from SGR 1935+2154 observed by GECAM-B, \emph{Fermi}-GBM, and \emph{INTEGRAL}/API-ACS \citep{2021ApJ...920...43X}. The half-width of the annulus region obtained by GECAM-B and \emph{Fermi}-GBM is 0.4 $^{\circ}$ (1$\sigma$).

\section{In-flight performance} \label{sec:IOP}

The two GECAM satellites were co-launched on 2020 December 10 (Beijing Time) \citep{2021arXiv211204772L}. GECAM is scheduled to work in a survey mode, where the GECAM points opposite to the Earth. Because of the power issue, GECAM-B was set to the ``pointing'' mode with the solar panel orienting towards the Sun since January 14, 2021, in order to provide the maximum energy to the spacecraft. Unfortunately, at the date of this writing, GECAM-A failed to turn on the payload, due to a power supply issue. GECAM-B works for about 10 hours per day.

\subsection{Trigger statistics and analysis} \label{sec:41}

\begin{figure*}
\centering
\includegraphics[scale=0.5]{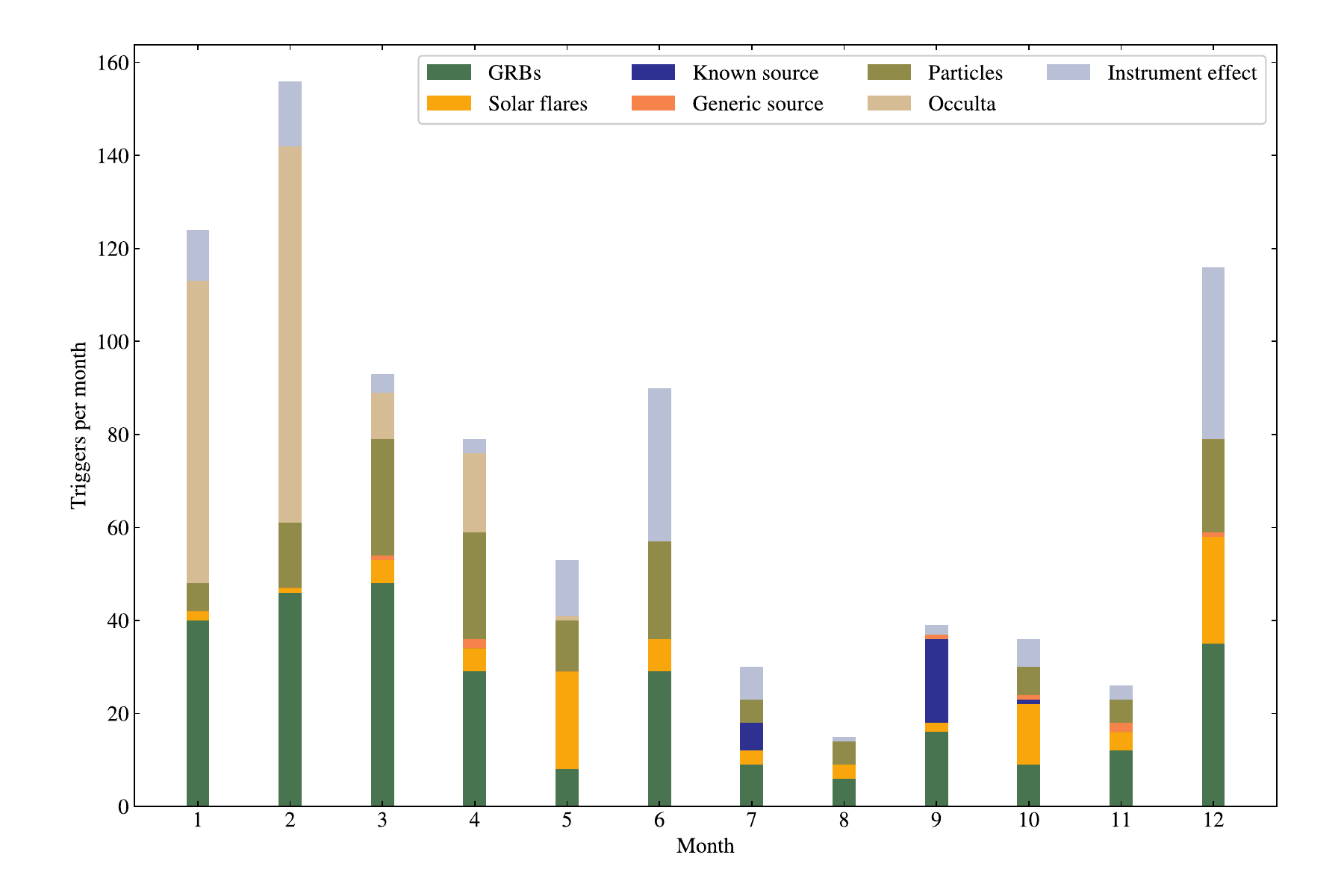}
\caption{Monthly trigger statistics for year 2021. The trigger classification reported here are the result of auto-ground analysis.}
\label{fig:cla_hist}
\end{figure*}

During its first year (2021) of operation, GECAM was triggered 858 times \footnote{GECAM was initially triggered 1029 times in flight. Due to the dropped data packets during the real-time communication stream, we received only 858 triggers. See section \ref{sec:42} for details.} on a variety of transient events in flight (see Figure \ref{fig:cla_hist}): 42 of these are verified as GRBs, 32 as bursts from SGRs, 1 as Type-I X-ray burst (XRB) from X-ray binary 4U 0614+09 \citep{2021arXiv211204790C1}, and 783 as others (solar flares, charged particles, earth occultation, or instrument effect) by Burst Advocate (BA). Table \ref{tab:grb_classify} shows the number of events classified by the GIRTLS, BAS, and BA. For example, 666 triggers are classified as GRBs by the GIRTLS. Among them, 42 are ``real'' GRBs, 32 are SGRs, 1 is an XRB, and 591 are Others. The GIRTLS has a 100\% success rate classifying GRBs, but only a 24\% success rate of not identifying other events as GRBs. Compared to GIRTLS, 288 triggers are classified as GRBs by the BAS, and 34 of these are ``real'' GRBs. Eight ``real'' GRBs were misclassified as Generic sources by the BAS, as they were located near the galactic plane. The BAS has an 80\% success rate classifying GRBs, and a 70\% success rate of not identifying other events as GRBs. Most of those mis-classified as GRBs are particle events and instrument effects. We will continue to investigate additional improvements to the classification algorithms.

\begin{table*}[]
    \caption{The number of events classified by GIRTLS, BAS, and BA. Others include solar flares, particles, occulta and instrument effect.}
    \label{tab:grb_classify}
    \centering
    \begin{tabular}{ccccccc}
    \hline\hline
     &  &   &  \multicolumn{4}{c}{Event Classified by BA} \\
    \cmidrule(r){4-7}
     & Classified As & & GRB & \multicolumn{2}{c}{Known Source} & Others \\
     \cmidrule(r){5-6}
    &  &  &  &  SGR  & XRB & \\
    \hline
    \hline
    \multirow{3}{*}{GIRTLS} & GRB & 666 & 42 & 32 & 1 & 591 \\
    & Known Source & 185 & - & - & - & 185 \\
    & Occult & 7 & - & - & - & 7 \\
    \hline
    \multirow{8}{*}{BAS} & GRB & 288 & 34 & 5 & - & 249 \\
     & Known Source & 25 & - & 25 & - & - \\
     & Generic Source & 8 & 8 & - & - & - \\
     & Solar Flare & 89 & - & - & - & 89 \\
     & Occulta & 174 & - & - & - & 174 \\
     & Particles & 141 & - & - & - & 141 \\
     & Instrument Effect & 133 & - & 2 & 1 & 130 \\
    \hline
    \end{tabular}
    \footnotesize{}
\end{table*}

\begin{figure*}
\centering
\includegraphics[scale=0.5]{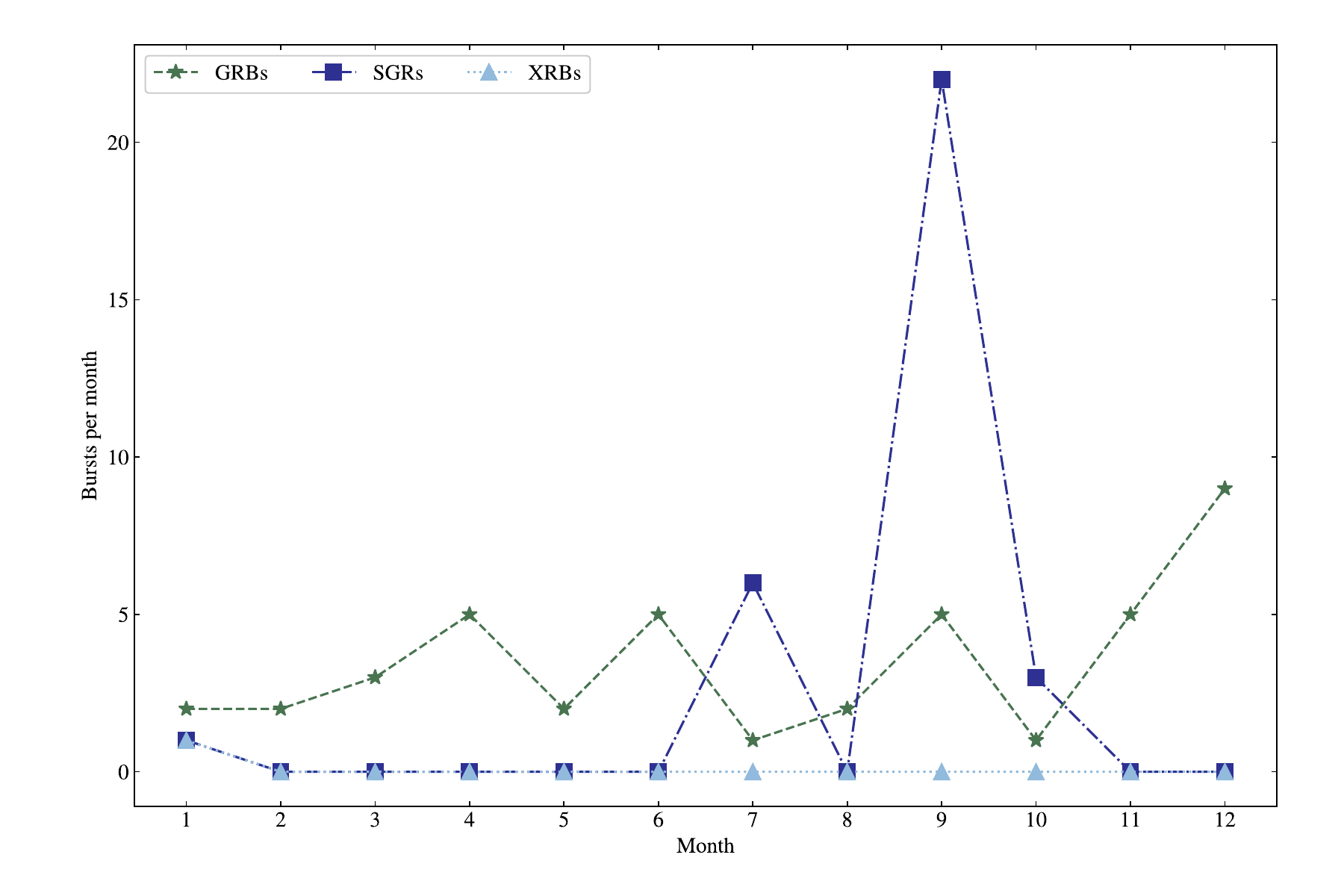}
\caption{Monthly bursts for year 2021. GECAM-B detected 42 GRBs (green star), 32 SGRs, 31 of them are from SGR 1935+2154 and the other one is from SGR J1555.2-5402 (blue square), and 1 XRB from 4U 0614+09 (light blue triangle) in 2021. }
\label{fig:cla_hist2}
\end{figure*}

The monthly trigger statistics over the first year of the mission is shown in Figure \ref{fig:cla_hist}. The higher rate of triggers in the beginning six months is due to the temperatures of the SiPM exceeded the design specifications ($-20\pm3 ^{\circ}$C) when the spacecraft adjusted its attitude mode. This leads to an increase of the thermal noise and may give false triggers. The SiPM is also prone to significantly increased thermal noise caused by on-orbit radiation damages, thereby decreasing its signal-to-noise ratio. This is clearly suggested by a significant decrease in the rate of detected triggers on the occulation of Sco X--1 after April, which has a soft spectra (see Figure \ref{fig:cla_hist}). Thus, we raised the low-energy threshold of GRD on December 30, 2020, January 5 and 18, 2021, and February 19, 2021, and the current low-energy threshold of GRD is about 15 keV. In addition, on Januray 27, 2021 we presented the first report of the reactivation of SGR J1935+2154 \citep{2021GCN.29363....1H}. GECAM also detected a series of bursts from this source in July and September of 2021.

\begin{figure*}
\centering
\includegraphics[scale=0.5]{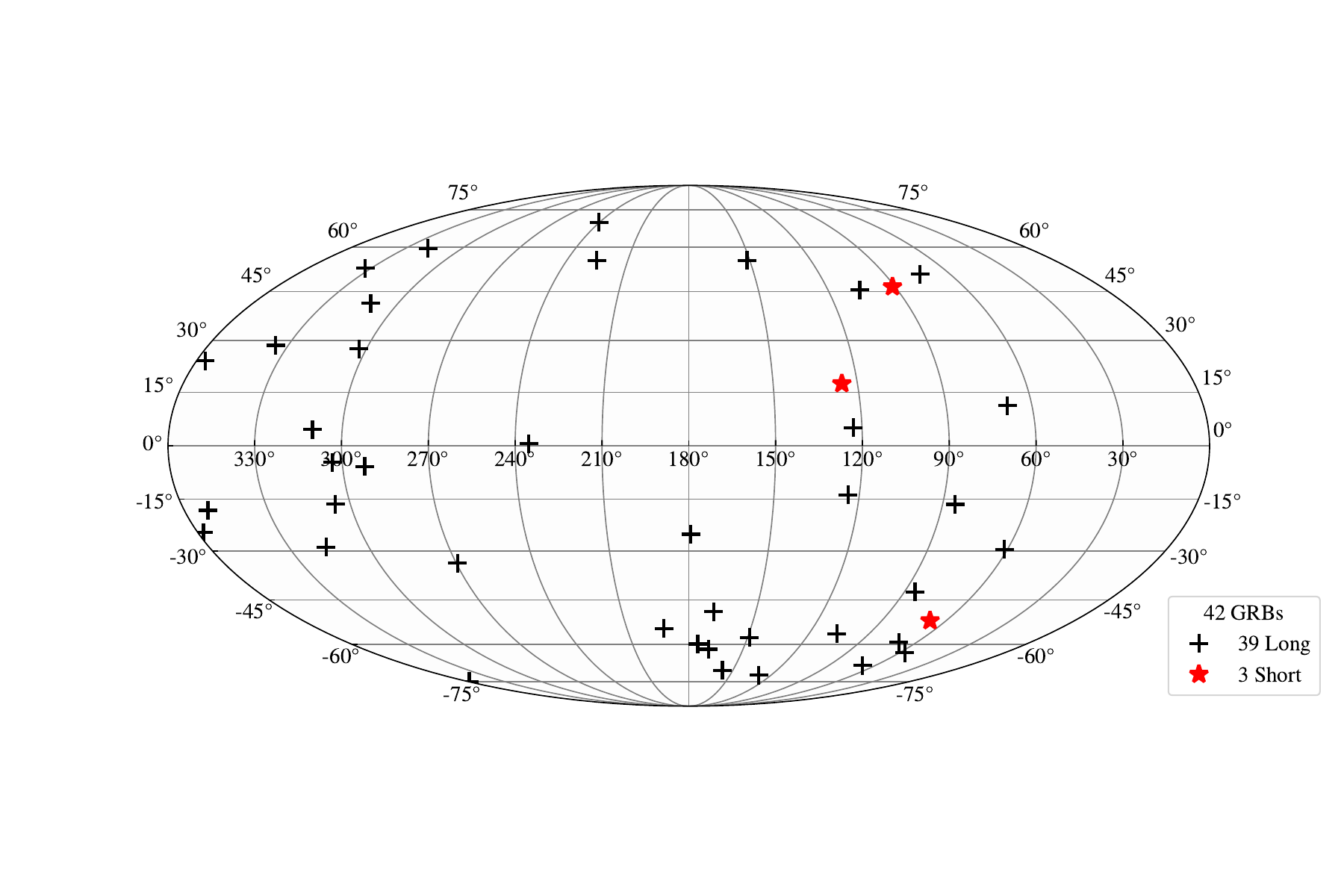}
\caption{Sky distribution of triggered GRBs by GECAM in-flight from 2021, in celestial coordinates. Crosses indicate long GRBs; asterisks indicate short GRBs. }
\label{fig:grb_loc}
\end{figure*}

\begin{figure}
\centering
\includegraphics[scale=0.8]{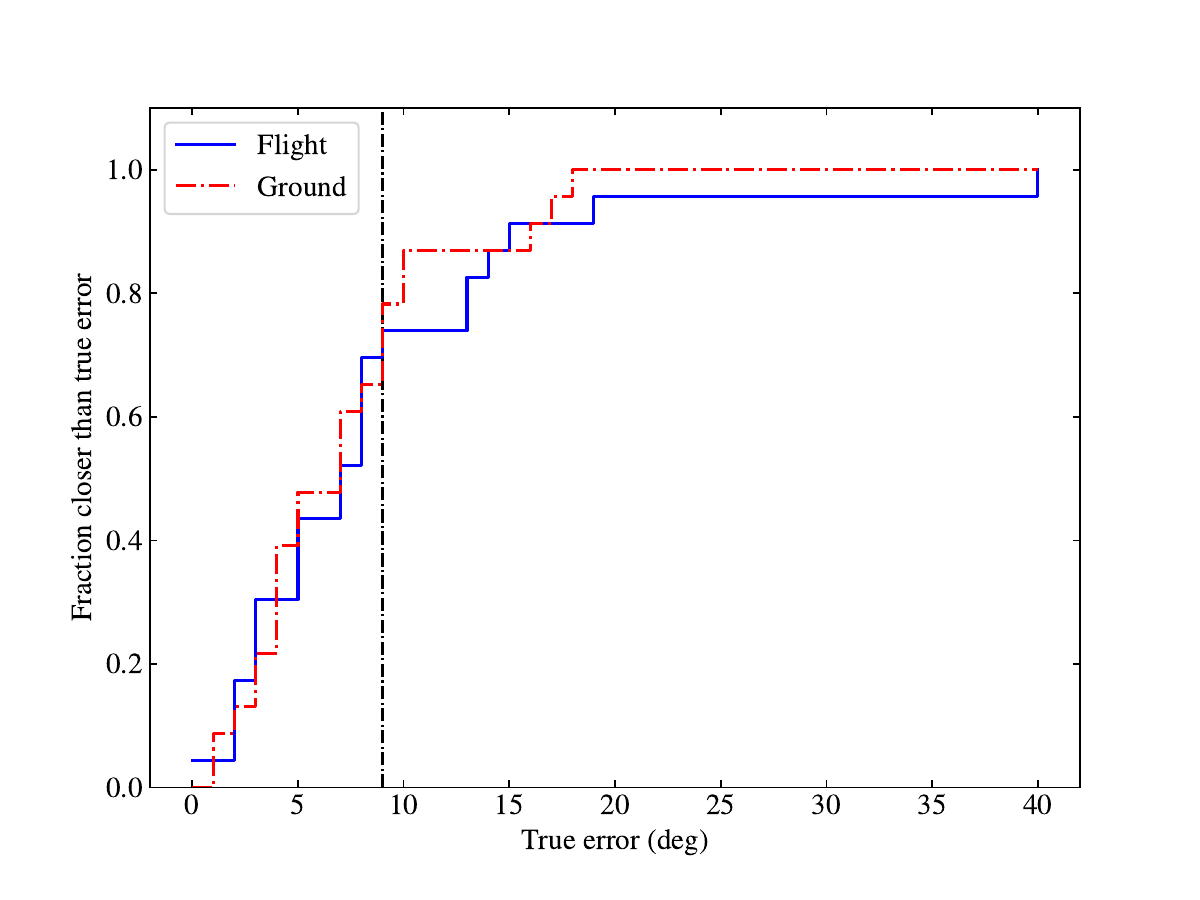}
\caption{The fraction of GECAM in-flight (42 GRBs) and ground (37 GRBs) localizations lying within a given offset from the real position. The vertical dot-dashed line indicates the 68\% containment radius.}
\label{fig:grb_loc_dist}
\end{figure}

Table \ref{tab:grb_cat} summarizes all 42 in-flight triggered GRBs from the first year's operation of GECAM. Figure \ref{fig:grb_loc} shows the sky distribution of the GRBs in celestial coordinates. There are 27 GRBs that are localized by other instruments (e.g., \emph{Swift}-BAT, \emph{Fermi}-GBM) or the IPN. These reference locations are also listed in Table \ref{tab:grb_cat}. Figure \ref{fig:grb_loc_dist} shows the fraction of GECAM in flight and ground localizations within a given offset from the reference location. The vertical dot-dashed line shows that 68\% of the reference locations are contained in a $\sim9^{\circ}$ region for both in-flight and ground locations.

\begin{table*}[]\tiny
    \caption{List of 42 GRBs triggered on-board of GECAM. }
    \label{tab:grb_cat}
    \centering
    {\tiny
    \begin{tabular}{ccccccccccccc}
    \hline\hline
    GRB Name & Trigger Time & Duration & HTR LC$^a$ & \multicolumn{3}{c}{Flight Location } & \multicolumn{3}{c}{BD Ground Location } & \multicolumn{2}{c}{Ref. Location} & Ref. Source \\
    \cmidrule(r){5-7}  \cmidrule(r){8-10} \cmidrule(r){11-12}
     & (UT) &  &  & Ra ($^{\circ}$)  & Dec ($^{\circ}$) & Err ($^{\circ}$) & Ra ($^{\circ}$) & Dec ($^{\circ}$) & Err ($^{\circ}$) & Ra ($^{\circ}$) & Dec ($^{\circ}$) &  \\
    \hline
    GRB 210120A & 07:10:48.550 & Long & N & 151.4 & 55.3 & 3.2 & 171.8 & 9.4 & 10.1 & 163.18 & 15.40 & MASTER \citep{2021GCN.29339....1L} \\
    GRB 210126A & 10:00:10.600 & Long & N & 106.1 & -56.3 & 6.1 & 112.9 & -53.3 & 5.2 & 90.2 & -66.0 & BALROG \citep{2021GCN.29354....1K} \\
    GRB 210204A & 06:30:00.600 & Long &  N & 123.0 & 5.1 & 3.3 & 121.3 & 8.3 & 3.2 & 117.08 & 11.41 & \emph{Swift}-XRT \citep{2021GCN.29413....1K} \\
    GRB 210228A & 06:38:32.600 & Long & N  & 85.6 & -42.5 & 1.8 & 85.5 & -41.8 & 2.1 & -$^b$ & - & - \\
    GRB 210307B & 05:56:39.100 & Short & N  & 125.6 & 17.5 & 7.3 & 131.8 & 13.2 & 15.9 & -$^b$ & - & - \\
    GRB 210317A & 09:08:28.550 & Long & N  & 157.1 & -70.1 & 2.4 & 153.2 & -68.1 & 2.8 & 154.2 & -64.9 & \emph{Fermi}-GBM \citep{2021GCN.29659....1F} \\
    GRB 210330A & 12:45:46.600 & Long &  N & 168.9 & -48.8 & 4.7 & 164.3 & -55.6 & 3.3 & -$^b$ & - & - \\
    GRB 210401A & 23:21:14.350 & Long &  N & 269.6 & -33.6 & 5.5 & 272.0 & -37.7 & 3.9 & 263.1 & -31.6 & \emph{Fermi}-GBM \citep{2021GCN.29739....1F} \\
    GRB 210409A & 21:28:27.950 & Long &  N & 69.4 & -59.3 & 2.8 & 70.9 & -58.3 & 3.0 & -$^b$ & - & - \\
    GRB 210413A & 01:07:25.600 & Long & Y & 68.6 & 11.3 & 3.6 & 63.6 & 8.4 & 7.4 & -$^b$ & - & - \\
    GRB 210425A & 07:07:04.200 & Short & Y & 67.5 & -51.8 & 2.7 & 66.8 & -50.1 & 12.3 & -$^b$ & - & - \\
    GRB 210427A & 04:57:13.100 & Long &  N & 175.3 & -59.9 & 5.8 & 160.1 & -69.1 & 1.8 & 177.79 & -52.87 & IPN \citep{2021GCN.29997....1H} \\
    GRB 210511B & 11:26:40.600 & Long &  N & 318.0 & 59.5 & 3.2 & 320.3 & 60.1 & 3.7 & 312.91 & 58.38 & IPN \citep{2021GCN.30002....1H} \\
    GRB 210520A & 19:07:03.550 & Long & N  & 129.0 & -72.0 & 5.6 & 126.5 & -72.9 & 5.8 & 123.0 & -69.4 & \emph{Fermi}-LAT \citep{2021GCN.30062....1D} \\
    GRB 210602B & 20:46:04.400 & Long & Y & 192.0 & -54.4 & 24.5 & 165.1 & -62.8 & 19.5 & -$^b$ & - & - \\
    GRB 210606B & 22:41:08.100 & Long & N  & 85.5 & -16.5 & 1.0 & 87.8 & -18.3 & 1.4 & 87.99 & -16.69 & IPN \citep{2021GCN.30154....1H} \\
    GRB 210619B & 00:00:00.950$^c$ & Long & N  & 334.7 & 28.6 & 31.0 & 318.8 & 29.2 & 7.4 & 319.71 & 33.85 & \emph{Swift}-XRT \citep{2021GCN.30261....1D} \\
    GRB 210622B & 10:33:02.600 & Long & Y & 123.9 & -13.8 & 4.8 & -$^d$ & - & - & 126.5 & -13.1 & \emph{Fermi}-GBM \citep{2021GCN.30298....1F} \\
    GRB 210627B & 17:57:21.550 & Long & N  & 235.3 & 0.6 & 3.6 & -$^d$ & - & - & 230.5 & -5.9 & BALROG \citep{2021GCN.30333....1B} \\
    GRB 210719A & 02:24:59.500 & Long & Y & 73.8 & 50.7 & 4.0 & 74.8 & 51.4 & 5.5 & -$^b$ & - & - \\
    GRB 210822A & 09:18:18.000 & Long &  N & 310.3 & 4.5 & 1.0 & 298.3 & 1.0 & 1.2 & 304.44 & 5.28 & \emph{Swift}-XRT \citep{2021GCN.30677....1P} \\
    GRB 210827B & 10:10:16.600 & Long & N  & 305.31 & -16.4 & 6.2 & 307.5 & -19.2 & 3.5 & 305.1 & -17.3 & BALROG \citep{2021GCN.30727....1B} \\
    GRB 210909B & 20:02:33.600 & Long & N  & 358.7 & -74.9 & 1.1 & -$^d$ & - & - & -$^b$ & - & - \\
    GRB 210919A & 00:28:33.800 & Short & Y$^e$ & 91.2 & 46.5 & 6.5 & -$^d$ & - & - & 80.25 & 1.31 & \emph{Swift}-XRT \citep{2021GCN.30850....1G} \\
    GRB 210925A & 19:12:34.600 & Long & Y & 357.8 & -24.6 & 3.5 & --$^d$ & -- & -- & -347.4 & -17.1 & \emph{Fermi}-GBM \citep{2021GCN.30880....1F} \\
    GRB 210926A & 20:52:28.250 & Long & N  & 351.5 & -18.2 & 2.9 & 351.9 & -23.6 & 7.1 & -$^b$ & - & - \\
    GRB 210927B & 23:54:45.600 & Long & Y & 240.3 & 69.5 & 9.9 & 249.6 & 70.4 & 3.3 & 263.02 & 73.77 & IPN \citep{2021GCN.30956....1K} \\
    GRB 211022A & 00:47:28.100 & Long &  N & 149.1 & -57.6 & 3.2 & 164.5 & -50.6 & 2.5 & 161.61 & -51.01 & IPN \citep{2021GCN.31024....1K} \\
    GRB 211102B & 14:05:35.350 & Long &  N & 303.4 & -4.6 & 2.8 & 298.1 & 2.9 & 2.4 & 315.4 & 1.5 & BALROG \citep{2021GCN.31033....1B} \\
    GRB 211105A & 04:35:20.200 & Long & N  & 68.5 & -68.0 & 2.6 & 81.9 & -57.4 & 1.3 & -$^b$ & - & - \\
    GRB 211109C & 07:51:02.200 & Long &  N & 169.1 & -61.8 & 8.2 & 183.6 & -55.8 & 5.6 & -$^b$ & - & - \\
    GRB 211110A & 03:26:29.600 & Long &  N & 61.2 & -29.5 & 4.2 & 64.5 & -30.3 & 5.4 & -$^b$ & - & - \\
    GRB 211120A & 23:05:20.600 & Long &  N & 311.3 & 41.3 & 1.0 & 305.0 & 41.4 & 1.3 & 315.14 & 42.86 & IPN \citep{2021GCN.31129....1K} \\
    GRB 211204C & 21:37:00.250 & Long &  N & 332.8 & 52.7 & 2.2 & 334.1 & 55.1 & 1.6 & 338.95 & 52.17 & IPN \citep{2021GCN.31177....1K} \\
    GRB 211211B & 21:48:43.050 & Long & N  & 225.2 & 55.3 & 3.5 & 225.8 & 52.2 & 3.8 & 228.1 & 48.4 & \emph{Fermi}-GBM \citep{2021GCN.31207....1F} \\
    GRB 211216A & 06:45:56.050 & Long & N  & 57.2 & -63.0 & 5.7 & 59.1 & -56.0 & 2.3 & 51.0 & -62.2 & BALROG \citep{2021GCN.31239....1B} \\
    GRB 211216B & 13:21:08.550 & Long &  N & 106.2 & 45.5 & 5.8 & 102.0 & 49.6 & 4.6 & 100.8 & 60.1 & \emph{Fermi}-GBM \citep{2021GCN.31244....1F} \\
    GRB 211217A & 07:04:30.550 & Long & Y & 356.8 & 24.1 & 8.2 & 5.2 & 26.9 & 10.9 & -$^b$ & - & - \\
    GRB 211223A & 02:41:18.900 & Long &  N & 316.0 & -28.9 & 5.6 & 320.2 & -30.7 & 4.4 & -$^b$ & - & - \\
    GRB 211229A & 03:30:05.200 & Long & Y  & 302.7 & 27.6 & 8.7 & 302.1 & 30.9 & 10.5 & 295.04 & 23.13 & \emph{Swift}-BAT \citep{2021GCN.31334....1T} \\
    GRB 211229B & 22:18:43.150 & Long &  N & 179.3 & -25.1 & 2.9 & 176.6 & -24.3 & 1.8 & 185.0 & -18.4 & \emph{Fermi}-GBM \citep{2021GCN.31336....1F} \\
    GRB 211231A & 07:00:35.050 & Long &  N & 292.3 & -5.8 & 2.7 & 289.6 & -16.4 & 2.7 & 292.2 & -24.9 & \emph{Fermi}-GBM \citep{2021GCN.31343....1F} \\
    \hline
    \end{tabular}
    }
    \footnotesize{$^a$ HTR LC: High temporal resolution light curve. There are 10 out of 42 GRBs had this light curve. \\
    $^b$ No other detection or location.\\
    $^c$ The bursts have multiple peaks. GECAM triggers at the latter peak. The trigger time of GECAM is 2021-06-20T00:00:00.950. \\
    $^d$ No ground location due to dropped data packets.\\
    $^e$ We do not received the high temporal resolution light curve from BeiDou due to the dropped data packets. }
\end{table*}

\subsection{BeiDou short message Performance} \label{sec:42}

The performance of the BeiDou-3 short message service is presented in this section. The time latency and the success rate of messages transmission are given.

\begin{figure}
\centering
\includegraphics[scale=1.]{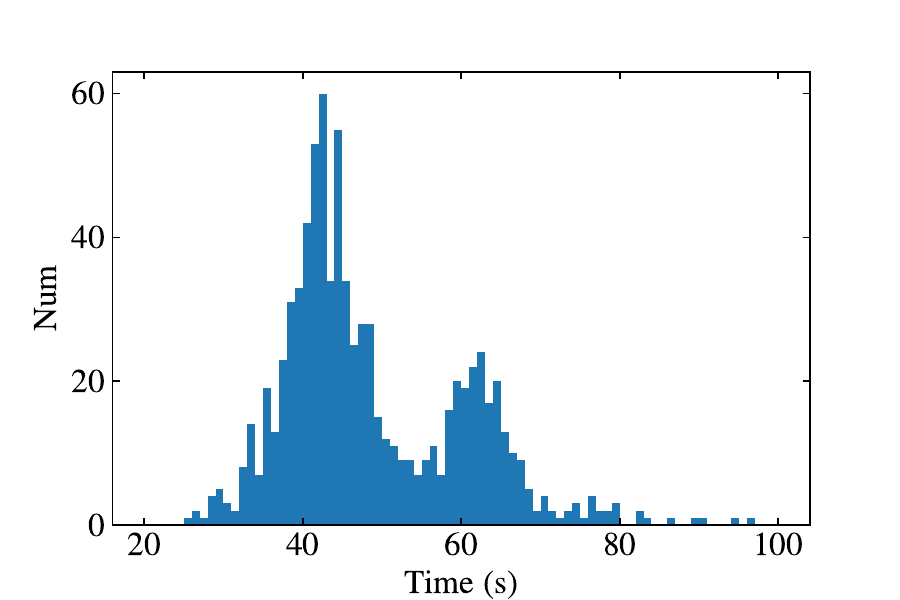}
\caption{Histogram of the time delays between the trigger time and the receiving time of the first or second short message of the trigger.}
\label{fig:trigger_delay}
\end{figure}

Figure \ref{fig:trigger_delay} shows the time delay between the trigger time and the receiving time of the first or second short message of the trigger. The average time delay is 45 s and the minimum time delay is 25 s. About 95\% of the triggers have time delays of less than 67 s. This is necessary for follow up observations and has led to several observations, e.g., \citet{2021GCN.31047....1D,2021GCN.31048....1L}. The time delay includes two parts. The first one is the delay from on-board signal processing. For short triggers, it takes $\sim$5 s to process the data, while for long triggers, it takes about 20 s. The second one is the delay between the message sending on-board and receiving on ground via the BeiDou short message service. Since the message is transmitted every 17 s, there will be an extra 17 s delay if the previous message failed to be received.

\begin{figure}
\centering
\includegraphics[scale=0.8]{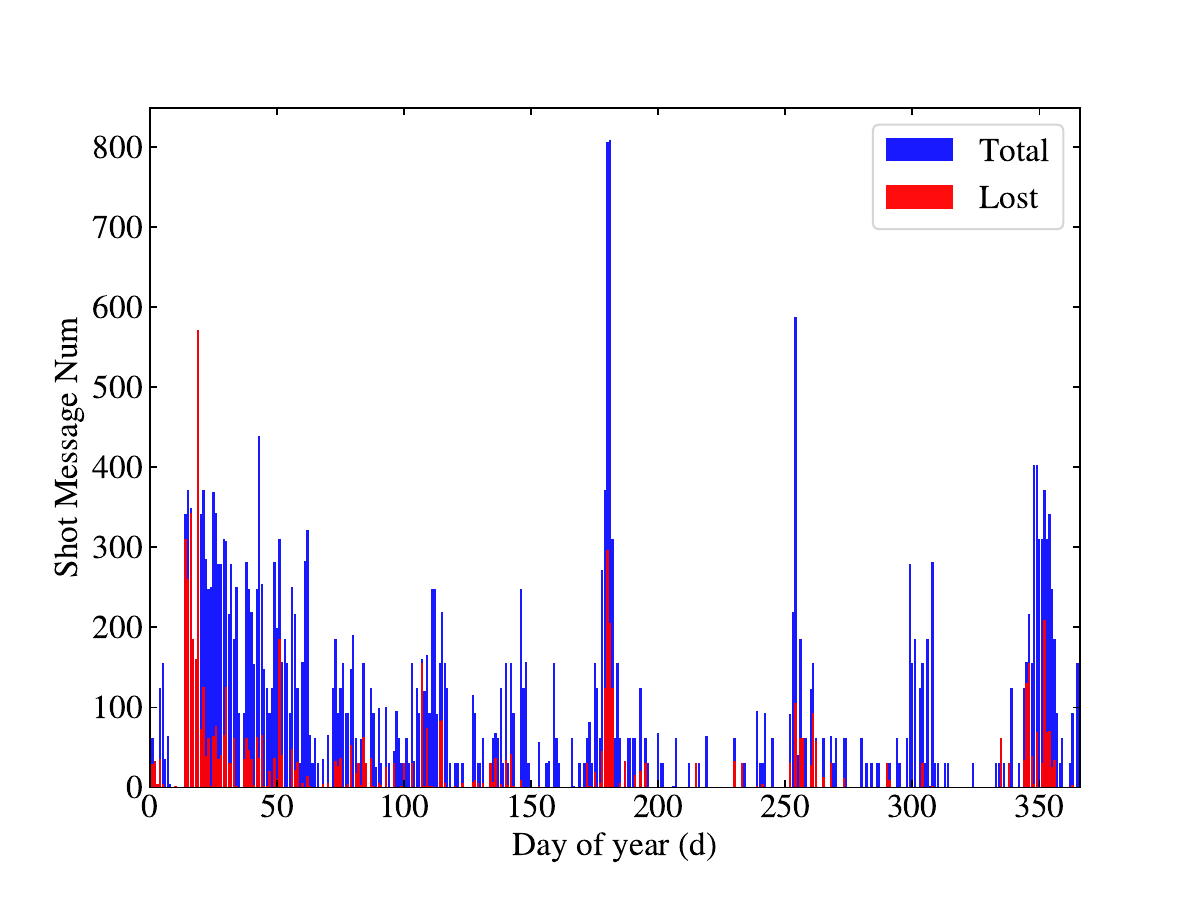}
\caption{Total number (blue) of messages and lost messages (red) per day.}
\label{fig:bdm_dr}
\end{figure}

The BeiDou short messages is not only transmitted in real-time, but also stored in the on-board storage and transmitted via the X-band ground station. We can thereby estimate the success rate of transmissions by comparing the data from the two methods. Figure \ref{fig:bdm_dr} shows the total number and the lost number of BeiDou short messages per day in 2021. On around January 15, most of the messages failed to transmit due to the attitude of the satellite. Because of the power supply issue, the satellite has to be frequently turn-off, which makes some messages fail to be timely sent before the satellite shutting down. Regardless of the satellite status, the success rate is 94.6\%, which is consistent with the official result given by the Beidou system which has a theoretical and test success rate of 95\% and 97.1\% \citep{2021AdSpR..67.1701L}.

\section{Conclusions and Perspectives} \label{sec:CS}

GECAM is the China's first transient explorer with a real-time alert system, who is capable of distributing GRB coordinates to ground observers within minutes using the BeiDou-3 short message service. During the first year of operation, GECAM had been triggered 858 times in flight, of which 42 are GRBs. The BAS processes the trigger alert data and provides refined classifications and localizations. The burst alert data can be transmitted to our collaborations within $\sim$1 minute. As of this writing, we are also collaborating with the GCN team on disseminating the notices via GCN. The in-flight performance shows that GECAM real-time BAS based on the BeiDou-3 short message service operates stably and efficiently. It has been applied to the subsequent GRB mission High Energy Burst Searcher (HEBS), which is a gamma-ray burst monitor on-board an experimental satellite to be launched in 2022 \citep{2022MNRAS.tmp..994X}.

GECAM mission aims to detect and localize GRBs associated with GW events. But the low luminosity and flux of GRB 170817A suggest that a population of short GRBs may be missed due to the lack of on-board triggers. In addition to the automated flight triggers, GECAM will also provide a targeted coherent search for GRBs associated with GW events, and search the sub-threshold short GRBs which can be used to search for low-significance GW signals. Moreover, a further dedicated effort is ongoing to improve the ground localization, classification, and automatic alerting procedure. GECAM is going to play a crucial role in the LIGO, Virgo, and KAGRA forthcoming fourth observing run (O4) to search for and characterize the EM counterparts of GW events. It is also necessary to fully exploit the scientific potential of neutrinos and fast radio bursts, since these events also require high-energy EM observations for identification and further study.


\begin{acknowledgements}

The GECAM (Huairou-1) mission is supported by the Strategic Priority Research Program on Space Science of the Chinese Academy of Sciences. The authors thank the support from the Strategic Priority Research Program on Space Science (Grant No. XDA15360000, XDA15360300, XDA15360102, XDA15052700) of the Chinese Academy of Sciences, the National Natural Science Foundation of China (Grant No. U2031205, 12133007), and the National Key R\&D Program of China (2021YFA0718500, 2022YFF0711404).


\end{acknowledgements}

\bibliography{ms2022-0155}{}

\begin{thebibliography}{}
\expandafter\ifx\csname natexlab\endcsname\relax\def\natexlab#1{#1}\fi
\providecommand{\url}[1]{\href{#1}{#1}}
\providecommand{\dodoi}[1]{doi:~\href{http://doi.org/#1}{\nolinkurl{#1}}}
\providecommand{\doeprint}[1]{\href{http://ascl.net/#1}{\nolinkurl{http://ascl.net/#1}}}
\providecommand{\doarXiv}[1]{\href{https://arxiv.org/abs/#1}{\nolinkurl{https://arxiv.org/abs/#1}}}

\bibitem[{{Abbott} {et~al.}(2016{\natexlab{a}}){Abbott}, {Abbott}, {Abbott},
  {Abernathy}, {Acernese}, {Ackley}, {Adams}, {Adams}, {Addesso}, {Adhikari},
  {Adya}, {Affeldt}, {Agathos}, {Agatsuma}, {Aggarwal}, {Aguiar}, {Aiello},
  {Ain}, {Ajith}, {Allen}, {Allocca}, {Altin}, {Anderson}, {Anderson}, {Arai},
  {Arain}, {Araya}, {Arceneaux}, {Areeda}, {Arnaud}, {Arun}, {Ascenzi},
  {Ashton}, {Ast}, {Aston}, {Astone}, {Aufmuth}, {Aulbert}, {Babak}, {Bacon},
  {Bader}, {Baker}, {Baldaccini}, {Ballardin}, {Ballmer}, {Barayoga},
  {Barclay}, {Barish}, {Barker}, {Barone}, {Barr}, {Barsotti}, {Barsuglia},
  {Barta}, {Bartlett}, {Barton}, {Bartos}, {Bassiri}, {Basti}, {Batch},
  {Baune}, {Bavigadda}, {Bazzan}, {Behnke}, {Bejger}, {Belczynski}, {Bell},
  {Bell}, {Berger}, {Bergman}, {Bergmann}, {Berry}, {Bersanetti}, {Bertolini},
  {Betzwieser}, {Bhagwat}, {Bhandare}, {Bilenko}, {Billingsley}, {Birch},
  {Birney}, {Birnholtz}, {Biscans}, {Bisht}, {Bitossi}, {Biwer}, {Bizouard},
  {Blackburn}, {Blair}, {Blair}, {Blair}, {Bloemen}, {Bock}, {Bodiya}, {Boer},
  {Bogaert}, {Bogan}, {Bohe}, {Bojtos}, {Bond}, {Bondu}, {Bonnand}, {Boom},
  {Bork}, {Boschi}, {Bose}, {Bouffanais}, {Bozzi}, {Bradaschia}, {Brady},
  {Braginsky}, {Branchesi}, {Brau}, {Briant}, {Brillet}, {Brinkmann},
  {Brisson}, {Brockill}, {Brooks}, {Brown}, {Brown}, {Brown}, {Buchanan},
  {Buikema}, {Bulik}, {Bulten}, {Buonanno}, {Buskulic}, {Buy}, {Byer},
  {Cabero}, {Cadonati}, {Cagnoli}, {Cahillane}, {Bustillo}, {Callister},
  {Calloni}, {Camp}, {Cannon}, {Cao}, {Capano}, {Capocasa}, {Carbognani},
  {Caride}, {Casanueva Diaz}, {Casentini}, {Caudill}, {Cavagli{\`a}},
  {Cavalier}, {Cavalieri}, {Cella}, {Cepeda}, {Baiardi}, {Cerretani},
  {Cesarini}, {Chakraborty}, {Chalermsongsak}, {Chamberlin}, {Chan}, {Chao},
  {Charlton}, {Chassande-Mottin}, {Chen}, {Chen}, {Cheng}, {Chincarini},
  {Chiummo}, {Cho}, {Cho}, {Chow}, {Christensen}, {Chu}, {Chua}, {Chung},
  {Ciani}, {Clara}, {Clark}, {Cleva}, {Coccia}, {Cohadon}, {Colla}, {Collette},
  {Cominsky}, {Constancio}, {Conte}, {Conti}, {Cook}, {Corbitt}, {Cornish},
  {Corsi}, {Cortese}, {Costa}, {Coughlin}, {Coughlin}, {Coulon}, {Countryman},
  {Couvares}, {Cowan}, {Coward}, {Cowart}, {Coyne}, {Coyne}, {Craig},
  {Creighton}, {Creighton}, {Cripe}, {Crowder}, {Cruise}, {Cumming},
  {Cunningham}, {Cuoco}, {Dal Canton}, {Danilishin}, {D'Antonio}, {Danzmann},
  {Darman}, {Da Silva Costa}, {Dattilo}, {Dave}, {Daveloza}, {Davier},
  {Davies}, {Daw}, {Day}, {De}, {DeBra}, {Debreczeni}, {Degallaix}, {De
  Laurentis}, {Del{\'e}glise}, {Del Pozzo}, {Denker}, {Dent}, {Dereli},
  {Dergachev}, {DeRosa}, {De Rosa}, {DeSalvo}, {Dhurandhar}, {D{\'\i}az}, {Di
  Fiore}, {Di Giovanni}, {Di Lieto}, {Di Pace}, {Di Palma}, {Di Virgilio},
  {Dojcinoski}, {Dolique}, {Donovan}, {Dooley}, {Doravari}, {Douglas},
  {Downes}, {Drago}, {Drever}, {Driggers}, {Du}, {Ducrot}, {Dwyer}, {Edo},
  {Edwards}, {Effler}, {Eggenstein}, {Ehrens}, {Eichholz}, {Eikenberry},
  {Engels}, {Essick}, {Etzel}, {Evans}, {Evans}, {Everett}, {Factourovich},
  {Fafone}, {Fair}, {Fairhurst}, {Fan}, {Fang}, {Farinon}, {Farr}, {Farr},
  {Favata}, {Fays}, {Fehrmann}, {Fejer}, {Feldbaum}, {Ferrante}, {Ferreira},
  {Ferrini}, {Fidecaro}, {Finn}, {Fiori}, {Fiorucci}, {Fisher}, {Flaminio},
  {Fletcher}, {Fong}, {Fournier}, {Franco}, {Frasca}, {Frasconi}, {Frede},
  {Frei}, {Freise}, {Frey}, {Frey}, {Fricke}, {Fritschel}, {Frolov}, {Fulda},
  {Fyffe}, {Gabbard}, {Gair}, {Gammaitoni}, {Gaonkar}, {Garufi}, {Gatto},
  {Gaur}, {Gehrels}, {Gemme}, {Gendre}, {Genin}, {Gennai}, {George}, {Gergely},
  {Germain}, {Ghosh}, {Ghosh}, {Ghosh}, {Giaime}, {Giardina}, {Giazotto},
  {Gill}, {Glaefke}, {Gleason}, {Goetz}, {Goetz}, {Gondan}, {Gonz{\'a}lez},
  {Castro}, {Gopakumar}, {Gordon}, {Gorodetsky}, {Gossan}, {Gosselin},
  {Gouaty}, {Graef}, {Graff}, {Granata}, {Grant}, {Gras}, {Gray}, {Greco},
  {Green}, {Greenhalgh}, {Groot}, {Grote}, {Grunewald}, {Guidi}, {Guo},
  {Gupta}, {Gupta}, {Gushwa}, {Gustafson}, {Gustafson}, {Hacker}, {Hall},
  {Hall}, {Hammond}, {Haney}, {Hanke}, {Hanks}, {Hanna}, {Hannam}, {Hanson},
  {Hardwick}, {Harms}, {Harry}, {Harry}, {Hart}, {Hartman}, {Haster},
  {Haughian}, {Healy}, {Heefner}, {Heidmann}, {Heintze}, {Heinzel}, {Heitmann},
  {Hello}, {Hemming}, {Hendry}, {Heng}, {Hennig}, {Heptonstall}, {Heurs},
  {Hild}, {Hoak}, {Hodge}, {Hofman}, {Hollitt}, {Holt}, {Holz}, {Hopkins},
  {Hosken}, {Hough}, {Houston}, {Howell}, {Hu}, {Huang}, {Huerta}, {Huet},
  {Hughey}, {Husa}, {Huttner}, {Huynh-Dinh}, {Idrisy}, {Indik}, {Ingram},
  {Inta}, {Isa}, {Isac}, {Isi}, {Islas}, {Isogai}, {Iyer}, {Izumi}, {Jacobson},
  {Jacqmin}, {Jang}, {Jani}, {Jaranowski}, {Jawahar}, {Jim{\'e}nez-Forteza},
  {Johnson}, {Johnson-McDaniel}, {Jones}, {Jones}, {Jonker}, {Ju}, {Haris},
  {Kalaghatgi}, {Kalogera}, {Kandhasamy}, {Kang}, {Kanner}, {Karki},
  {Kasprzack}, {Katsavounidis}, {Katzman}, {Kaufer}, {Kaur}, {Kawabe},
  {Kawazoe}, {K{\'e}f{\'e}lian}, {Kehl}, {Keitel}, {Kelley}, {Kells},
  {Kennedy}, {Keppel}, {Key}, {Khalaidovski}, {Khalili}, {Khan}, {Khan},
  {Khan}, {Khazanov}, {Kijbunchoo}, {Kim}, {Kim}, {Kim}, {Kim}, {Kim}, {Kim},
  {King}, {King}, {Kinzel}, {Kissel}, {Kleybolte}, {Klimenko}, {Koehlenbeck},
  {Kokeyama}, {Koley}, {Kondrashov}, {Kontos}, {Koranda}, {Korobko}, {Korth},
  {Kowalska}, {Kozak}, {Kringel}, {Krishnan}, {Kr{\'o}lak}, {Krueger}, {Kuehn},
  {Kumar}, {Kumar}, {Kuo}, {Kutynia}, {Kwee}, {Lackey}, {Landry}, {Lange},
  {Lantz}, {Lasky}, {Lazzarini}, {Lazzaro}, {Leaci}, {Leavey}, {Lebigot},
  {Lee}, {Lee}, {Lee}, {Lee}, {Lenon}, {Leonardi}, {Leong}, {Leroy},
  {Letendre}, {Levin}, {Levine}, {Li}, {Libson}, {Littenberg}, {Lockerbie},
  {Logue}, {Lombardi}, {London}, {Lord}, {Lorenzini}, {Loriette}, {Lormand},
  {Losurdo}, {Lough}, {Lousto}, {Lovelace}, {L{\"u}ck}, {Lundgren}, {Luo},
  {Lynch}, {Ma}, {MacDonald}, {Machenschalk}, {MacInnis}, {Macleod},
  {Maga{\~n}a-Sandoval}, {Magee}, {Mageswaran}, {Majorana}, {Maksimovic},
  {Malvezzi}, {Man}, {Mandel}, {Mandic}, {Mangano}, {Mansell}, {Manske},
  {Mantovani}, {Marchesoni}, {Marion}, {M{\'a}rka}, {M{\'a}rka}, {Markosyan},
  {Maros}, {Martelli}, {Martellini}, {Martin}, {Martin}, {Martynov}, {Marx},
  {Mason}, {Masserot}, {Massinger}, {Masso-Reid}, {Matichard}, {Matone},
  {Mavalvala}, {Mazumder}, {Mazzolo}, {McCarthy}, {McClelland}, {McCormick},
  {McGuire}, {McIntyre}, {McIver}, {McManus}, {McWilliams}, {Meacher},
  {Meadors}, {Meidam}, {Melatos}, {Mendell}, {Mendoza-Gandara}, {Mercer},
  {Merilh}, {Merzougui}, {Meshkov}, {Messenger}, {Messick}, {Meyers},
  {Mezzani}, {Miao}, {Michel}, {Middleton}, {Mikhailov}, {Milano}, {Miller},
  {Millhouse}, {Minenkov}, {Ming}, {Mirshekari}, {Mishra}, {Mitra},
  {Mitrofanov}, {Mitselmakher}, {Mittleman}, {Moggi}, {Mohan}, {Mohapatra},
  {Montani}, {Moore}, {Moore}, {Moraru}, {Moreno}, {Morriss}, {Mossavi},
  {Mours}, {Mow-Lowry}, {Mueller}, {Mueller}, {Muir}, {Mukherjee}, {Mukherjee},
  {Mukherjee}, {Mukund}, {Mullavey}, {Munch}, {Murphy}, {Murray}, {Mytidis},
  {Nardecchia}, {Naticchioni}, {Nayak}, {Necula}, {Nedkova}, {Nelemans},
  {Neri}, {Neunzert}, {Newton}, {Nguyen}, {Nielsen}, {Nissanke}, {Nitz},
  {Nocera}, {Nolting}, {Normandin}, {Nuttall}, {Oberling}, {Ochsner}, {O'Dell},
  {Oelker}, {Ogin}, {Oh}, {Oh}, {Ohme}, {Oliver}, {Oppermann}, {Oram},
  {O'Reilly}, {O'Shaughnessy}, {Ott}, {Ottaway}, {Ottens}, {Overmier}, {Owen},
  {Pai}, {Pai}, {Palamos}, {Palashov}, {Palomba}, {Pal-Singh}, {Pan}, {Pan},
  {Pankow}, {Pannarale}, {Pant}, {Paoletti}, {Paoli}, {Papa}, {Paris},
  {Parker}, {Pascucci}, {Pasqualetti}, {Passaquieti}, {Passuello},
  {Patricelli}, {Patrick}, {Pearlstone}, {Pedraza}, {Pedurand}, {Pekowsky},
  {Pele}, {Penn}, {Perreca}, {Pfeiffer}, {Phelps}, {Piccinni}, {Pichot},
  {Pickenpack}, {Piergiovanni}, {Pierro}, {Pillant}, {Pinard}, {Pinto},
  {Pitkin}, {Poeld}, {Poggiani}, {Popolizio}, {Post}, {Powell}, {Prasad},
  {Predoi}, {Premachandra}, {Prestegard}, {Price}, {Prijatelj}, {Principe},
  {Privitera}, {Prix}, {Prodi}, {Prokhorov}, {Puncken}, {Punturo}, {Puppo},
  {P{\"u}rrer}, {Qi}, {Qin}, {Quetschke}, {Quintero}, {Quitzow-James}, {Raab},
  {Rabeling}, {Radkins}, {Raffai}, {Raja}, {Rakhmanov}, {Ramet}, {Rapagnani},
  {Raymond}, {Razzano}, {Re}, {Read}, {Reed}, {Regimbau}, {Rei}, {Reid},
  {Reitze}, {Rew}, {Reyes}, {Ricci}, {Riles}, {Robertson}, {Robie}, {Robinet},
  {Rocchi}, {Rolland}, {Rollins}, {Roma}, {Romano}, {Romano}, {Romanov},
  {Romie}, {Rosi{\'n}ska}, {Rowan}, {R{\"u}diger}, {Ruggi}, {Ryan}, {Sachdev},
  {Sadecki}, {Sadeghian}, {Salconi}, {Saleem}, {Salemi}, {Samajdar}, {Sammut},
  {Sampson}, {Sanchez}, {Sandberg}, {Sandeen}, {Sanders}, {Sanders},
  {Sassolas}, {Sathyaprakash}, {Saulson}, {Sauter}, {Savage}, {Sawadsky},
  {Schale}, {Schilling}, {Schmidt}, {Schmidt}, {Schnabel}, {Schofield},
  {Sch{\"o}nbeck}, {Schreiber}, {Schuette}, {Schutz}, {Scott}, {Scott},
  {Sellers}, {Sengupta}, {Sentenac}, {Sequino}, {Sergeev}, {Serna},
  {Setyawati}, {Sevigny}, {Shaddock}, {Shaffer}, {Shah}, {Shahriar}, {Shaltev},
  {Shao}, {Shapiro}, {Shawhan}, {Sheperd}, {Shoemaker}, {Shoemaker}, {Siellez},
  {Siemens}, {Sigg}, {Silva}, {Simakov}, {Singer}, {Singer}, {Singh}, {Singh},
  {Singhal}, {Sintes}, {Slagmolen}, {Smith}, {Smith}, {Smith}, {Smith}, {Son},
  {Sorazu}, {Sorrentino}, {Souradeep}, {Srivastava}, {Staley}, {Steinke},
  {Steinlechner}, {Steinlechner}, {Steinmeyer}, {Stephens}, {Stevenson},
  {Stone}, {Strain}, {Straniero}, {Stratta}, {Strauss}, {Strigin}, {Sturani},
  {Stuver}, {Summerscales}, {Sun}, {Sutton}, {Swinkels}, {Szczepa{\'n}czyk},
  {Tacca}, {Talukder}, {Tanner}, {T{\'a}pai}, {Tarabrin}, {Taracchini},
  {Taylor}, {Theeg}, {Thirugnanasambandam}, {Thomas}, {Thomas}, {Thomas},
  {Thorne}, {Thorne}, {Thrane}, {Tiwari}, {Tiwari}, {Tokmakov}, {Tomlinson},
  {Tonelli}, {Torres}, {Torrie}, {T{\"o}yr{\"a}}, {Travasso}, {Traylor},
  {Trifir{\`o}}, {Tringali}, {Trozzo}, {Tse}, {Turconi}, {Tuyenbayev},
  {Ugolini}, {Unnikrishnan}, {Urban}, {Usman}, {Vahlbruch}, {Vajente},
  {Valdes}, {Vallisneri}, {van Bakel}, {van Beuzekom}, {van den Brand}, {Van
  Den Broeck}, {Vander-Hyde}, {van der Schaaf}, {van Heijningen}, {van Veggel},
  {Vardaro}, {Vass}, {Vas{\'u}th}, {Vaulin}, {Vecchio}, {Vedovato}, {Veitch},
  {Veitch}, {Venkateswara}, {Verkindt}, {Vetrano}, {Vicer{\'e}}, {Vinciguerra},
  {Vine}, {Vinet}, {Vitale}, {Vo}, {Vocca}, {Vorvick}, {Voss}, {Vousden},
  {Vyatchanin}, {Wade}, {Wade}, {Wade}, {Waldman}, {Walker}, {Wallace},
  {Walsh}, {Wang}, {Wang}, {Wang}, {Wang}, {Wang}, {Ward}, {Ward}, {Warner},
  {Was}, {Weaver}, {Wei}, {Weinert}, {Weinstein}, {Weiss}, {Welborn}, {Wen},
  {We{\ss}els}, {Westphal}, {Wette}, {Whelan}, {Whitcomb}, {White}, {Whiting},
  {Wiesner}, {Wilkinson}, {Willems}, {Williams}, {Williams}, {Williamson},
  {Willis}, {Willke}, {Wimmer}, {Winkelmann}, {Winkler}, {Wipf}, {Wiseman},
  {Wittel}, {Woan}, {Worden}, {Wright}, {Wu}, {Yablon}, {Yakushin}, {Yam},
  {Yamamoto}, {Yancey}, {Yap}, {Yu}, {Yvert}, {Zadro{\.Z}ny}, {Zangrando},
  {Zanolin}, {Zendri}, {Zevin}, {Zhang}, {Zhang}, {Zhang}, {Zhang}, {Zhao},
  {Zhou}, {Zhou}, {Zhu}, {Zucker}, {Zuraw}, {Zweizig}, {LIGO Scientific
  Collaboration}, \& {Virgo Collaboration}}]{2016PhRvL.116f1102A}
{Abbott}, B.~P., {Abbott}, R., {Abbott}, T.~D., {et~al.} 2016{\natexlab{a}},
  \prl, 116, 061102, \dodoi{10.1103/PhysRevLett.116.061102}

\bibitem[{{Abbott} {et~al.}(2016{\natexlab{b}}){Abbott}, {Abbott}, {Abbott},
  {Abernathy}, {Acernese}, {Ackley}, {Adams}, {Adams}, {Addesso}, {Adhikari},
  {Adya}, {Affeldt}, {Agathos}, {Agatsuma}, {Aggarwal}, {Aguiar}, {Aiello},
  {Ain}, {Ajith}, {Allen}, {Allocca}, {Altin}, {Anderson}, {Anderson}, {Arai},
  {Araya}, {Arceneaux}, {Areeda}, {Arnaud}, {Arun}, {Ascenzi}, {Ashton}, {Ast},
  {Aston}, {Astone}, {Aufmuth}, {Aulbert}, {Babak}, {Bacon}, {Bader}, {Baker},
  {Baldaccini}, {Ballardin}, {Ballmer}, {Barayoga}, {Barclay}, {Barish},
  {Barker}, {Barone}, {Barr}, {Barsotti}, {Barsuglia}, {Barta}, {Barthelmy},
  {Bartlett}, {Bartos}, {Bassiri}, {Basti}, {Batch}, {Baune}, {Bavigadda},
  {Bazzan}, {Behnke}, {Bejger}, {Bell}, {Bell}, {Berger}, {Bergman},
  {Bergmann}, {Berry}, {Bersanetti}, {Bertolini}, {Betzwieser}, {Bhagwat},
  {Bhandare}, {Bilenko}, {Billingsley}, {Birch}, {Birney}, {Biscans}, {Bisht},
  {Bitossi}, {Biwer}, {Bizouard}, {Blackburn}, {Blair}, {Blair}, {Blair},
  {Bloemen}, {Bock}, {Bodiya}, {Boer}, {Bogaert}, {Bogan}, {Bohe}, {Bojtos},
  {Bond}, {Bondu}, {Bonnand}, {Boom}, {Bork}, {Boschi}, {Bose}, {Bouffanais},
  {Bozzi}, {Bradaschia}, {Brady}, {Braginsky}, {Branchesi}, {Brau}, {Briant},
  {Brillet}, {Brinkmann}, {Brisson}, {Brockill}, {Brooks}, {Brown}, {Brown},
  {Brown}, {Buchanan}, {Buikema}, {Bulik}, {Bulten}, {Buonanno}, {Buskulic},
  {Buy}, {Byer}, {Cadonati}, {Cagnoli}, {Cahillane}, {Bustillo}, {Callister},
  {Calloni}, {Camp}, {Cannon}, {Cao}, {Capano}, {Capocasa}, {Carbognani},
  {Caride}, {Diaz}, {Casentini}, {Caudill}, {Cavagli{\'a}}, {Cavalier},
  {Cavalieri}, {Cella}, {Cepeda}, {Baiardi}, {Cerretani}, {Cesarini},
  {Chakraborty}, {Chalermsongsak}, {Chamberlin}, {Chan}, {Chao}, {Charlton},
  {Chassande-Mottin}, {Chen}, {Chen}, {Cheng}, {Chincarini}, {Chiummo}, {Cho},
  {Cho}, {Chow}, {Christensen}, {Chu}, {Chua}, {Chung}, {Ciani}, {Clara},
  {Clark}, {Cleva}, {Coccia}, {Cohadon}, {Colla}, {Collette}, {Cominsky},
  {Constancio}, {Conte}, {Conti}, {Cook}, {Corbitt}, {Cornish}, {Corsi},
  {Cortese}, {Costa}, {Coughlin}, {Coughlin}, {Coulon}, {Countryman},
  {Couvares}, {Cowan}, {Coward}, {Cowart}, {Coyne}, {Coyne}, {Craig},
  {Creighton}, {Cripe}, {Crowder}, {Cumming}, {Cunningham}, {Cuoco}, {Dal
  Canton}, {Danilishin}, {D'Antonio}, {Danzmann}, {Darman}, {Dattilo}, {Dave},
  {Daveloza}, {Davier}, {Davies}, {Daw}, {Day}, {DeBra}, {Debreczeni},
  {Degallaix}, {De Laurentis}, {Del{\'e}glise}, {Del Pozzo}, {Denker}, {Dent},
  {Dereli}, {Dergachev}, {DeRosa}, {De Rosa}, {DeSalvo}, {Dhurandhar},
  {D{\'\i}az}, {Di Fiore}, {Di Giovanni}, {Di Lieto}, {Di Pace}, {Di Palma},
  {Di Virgilio}, {Dojcinoski}, {Dolique}, {Donovan}, {Dooley}, {Doravari},
  {Douglas}, {Downes}, {Drago}, {Drever}, {Driggers}, {Du}, {Ducrot}, {Dwyer},
  {Edo}, {Edwards}, {Effler}, {Eggenstein}, {Ehrens}, {Eichholz}, {Eikenberry},
  {Engels}, {Essick}, {Etzel}, {Evans}, {Evans}, {Everett}, {Factourovich},
  {Fafone}, {Fair}, {Fairhurst}, {Fan}, {Fang}, {Farinon}, {Farr}, {Farr},
  {Favata}, {Fays}, {Fehrmann}, {Fejer}, {Ferrante}, {Ferreira}, {Ferrini},
  {Fidecaro}, {Fiori}, {Fiorucci}, {Fisher}, {Flaminio}, {Fletcher},
  {Fournier}, {Franco}, {Frasca}, {Frasconi}, {Frei}, {Freise}, {Frey}, {Frey},
  {Fricke}, {Fritschel}, {Frolov}, {Fulda}, {Fyffe}, {Gabbard}, {Gair},
  {Gammaitoni}, {Gaonkar}, {Garufi}, {Gatto}, {Gaur}, {Gehrels}, {Gemme},
  {Gendre}, {Genin}, {Gennai}, {George}, {Gergely}, {Germain}, {Ghosh},
  {Ghosh}, {Giaime}, {Giardina}, {Giazotto}, {Gill}, {Glaefke}, {Goetz},
  {Goetz}, {Gondan}, {Gonz{\'a}lez}, {Castro}, {Gopakumar}, {Gordon},
  {Gorodetsky}, {Gossan}, {Gosselin}, {Gouaty}, {Graef}, {Graff}, {Granata},
  {Grant}, {Gras}, {Gray}, {Greco}, {Green}, {Groot}, {Grote}, {Grunewald},
  {Guidi}, {Guo}, {Gupta}, {Gupta}, {Gushwa}, {Gustafson}, {Gustafson},
  {Hacker}, {Hall}, {Hall}, {Hammond}, {Haney}, {Hanke}, {Hanks}, {Hanna},
  {Hannam}, {Hanson}, {Hardwick}, {Haris}, {Harms}, {Harry}, {Harry}, {Hart},
  {Hartman}, {Haster}, {Haughian}, {Heidmann}, {Heintze}, {Heitmann}, {Hello},
  {Hemming}, {Hendry}, {Heng}, {Hennig}, {Heptonstall}, {Heurs}, {Hild},
  {Hoak}, {Hodge}, {Hofman}, {Hollitt}, {Holt}, {Holz}, {Hopkins}, {Hosken},
  {Hough}, {Houston}, {Howell}, {Hu}, {Huang}, {Huerta}, {Huet}, {Hughey},
  {Husa}, {Huttner}, {Huynh-Dinh}, {Idrisy}, {Indik}, {Ingram}, {Inta}, {Isa},
  {Isac}, {Isi}, {Islas}, {Isogai}, {Iyer}, {Izumi}, {Jacqmin}, {Jang}, {Jani},
  {Jaranowski}, {Jawahar}, {Jim{\'e}nez-Forteza}, {Johnson}, {Jones}, {Jones},
  {Jonker}, {Ju}, {Kalaghatgi}, {Kalogera}, {Kandhasamy}, {Kang}, {Kanner},
  {Karki}, {Kasprzack}, {Katsavounidis}, {Katzman}, {Kaufer}, {Kaur}, {Kawabe},
  {Kawazoe}, {K{\'e}f{\'e}lian}, {Kehl}, {Keitel}, {Kelley}, {Kells},
  {Kennedy}, {Key}, {Khalaidovski}, {Khalili}, {Khan}, {Khan}, {Khan},
  {Khazanov}, {Kijbunchoo}, {Kim}, {Kim}, {Kim}, {Kim}, {Kim}, {Kim}, {King},
  {King}, {Kinzel}, {Kissel}, {Kleybolte}, {Klimenko}, {Koehlenbeck},
  {Kokeyama}, {Koley}, {Kondrashov}, {Kontos}, {Korobko}, {Korth}, {Kowalska},
  {Kozak}, {Kringel}, {Kr{\'o}lak}, {Krueger}, {Kuehn}, {Kumar}, {Kuo},
  {Kutynia}, {Lackey}, {Landry}, {Lange}, {Lantz}, {Lasky}, {Lazzarini},
  {Lazzaro}, {Leaci}, {Leavey}, {Lebigot}, {Lee}, {Lee}, {Lee}, {Lee}, {Lenon},
  {Leonardi}, {Leong}, {Leroy}, {Letendre}, {Levin}, {Levine}, {Li}, {Libson},
  {Littenberg}, {Lockerbie}, {Logue}, {Lombardi}, {Lord}, {Lorenzini},
  {Loriette}, {Lormand}, {Losurdo}, {Lough}, {L{\"u}ck}, {Lundgren}, {Luo},
  {Lynch}, {Ma}, {MacDonald}, {Machenschalk}, {MacInnis}, {Macleod},
  {Maga{\~n}a-Sandoval}, {Magee}, {Mageswaran}, {Majorana}, {Maksimovic},
  {Malvezzi}, {Man}, {Mandel}, {Mandic}, {Mangano}, {Mansell}, {Manske},
  {Mantovani}, {Marchesoni}, {Marion}, {M{\'a}rka}, {M{\'a}rka}, {Markosyan},
  {Maros}, {Martelli}, {Martellini}, {Martin}, {Martin}, {Martynov}, {Marx},
  {Mason}, {Masserot}, {Massinger}, {Masso-Reid}, {Matichard}, {Matone},
  {Mavalvala}, {Mazumder}, {Mazzolo}, {McCarthy}, {McClelland}, {McCormick},
  {McGuire}, {McIntyre}, {McIver}, {McManus}, {McWilliams}, {Meacher},
  {Meadors}, {Meidam}, {Melatos}, {Mendell}, {Mendoza-Gandara}, {Mercer},
  {Merilh}, {Merzougui}, {Meshkov}, {Messenger}, {Messick}, {Meyers},
  {Mezzani}, {Miao}, {Michel}, {Middleton}, {Mikhailov}, {Milano}, {Miller},
  {Millhouse}, {Minenkov}, {Ming}, {Mirshekari}, {Mishra}, {Mitra},
  {Mitrofanov}, {Mitselmakher}, {Mittleman}, {Moggi}, {Mohan}, {Mohapatra},
  {Montani}, {Moore}, {Moore}, {Moraru}, {Moreno}, {Morriss}, {Mossavi},
  {Mours}, {Mow-Lowry}, {Mueller}, {Mueller}, {Muir}, {Mukherjee}, {Mukherjee},
  {Mukherjee}, {Mukund}, {Mullavey}, {Munch}, {Murphy}, {Murray}, {Mytidis},
  {Nardecchia}, {Naticchioni}, {Nayak}, {Necula}, {Nedkova}, {Nelemans},
  {Neri}, {Neunzert}, {Newton}, {Nguyen}, {Nielsen}, {Nissanke}, {Nitz},
  {Nocera}, {Nolting}, {Normandin}, {Nuttall}, {Oberling}, {Ochsner}, {O'Dell},
  {Oelker}, {Ogin}, {Oh}, {Oh}, {Ohme}, {Oliver}, {Oppermann}, {Oram},
  {O'Reilly}, {O'Shaughnessy}, {Ottaway}, {Ottens}, {Overmier}, {Owen}, {Pai},
  {Pai}, {Palamos}, {Palashov}, {Palliyaguru}, {Palomba}, {Pal-Singh}, {Pan},
  {Pankow}, {Pannarale}, {Pant}, {Paoletti}, {Paoli}, {Papa}, {Paris},
  {Parker}, {Pascucci}, {Pasqualetti}, {Passaquieti}, {Passuello},
  {Patricelli}, {Patrick}, {Pearlstone}, {Pedraza}, {Pedurand}, {Pekowsky},
  {Pele}, {Penn}, {Perreca}, {Phelps}, {Piccinni}, {Pichot}, {Piergiovanni},
  {Pierro}, {Pillant}, {Pinard}, {Pinto}, {Pitkin}, {Poggiani}, {Popolizio},
  {Post}, {Powell}, {Prasad}, {Predoi}, {Premachandra}, {Prestegard}, {Price},
  {Prijatelj}, {Principe}, {Privitera}, {Prodi}, {Prokhorov}, {Puncken},
  {Punturo}, {Puppo}, {P{\"u}rrer}, {Qi}, {Qin}, {Quetschke}, {Quintero},
  {Quitzow-James}, {Raab}, {Rabeling}, {Radkins}, {Raffai}, {Raja},
  {Rakhmanov}, {Rapagnani}, {Raymond}, {Razzano}, {Re}, {Read}, {Reed},
  {Regimbau}, {Rei}, {Reid}, {Reitze}, {Rew}, {Reyes}, {Ricci}, {Riles},
  {Robertson}, {Robie}, {Robinet}, {Rocchi}, {Rolland}, {Rollins}, {Roma},
  {Romano}, {Romanov}, {Romie}, {Rosi{\'n}ska}, {Rowan}, {R{\"u}diger},
  {Ruggi}, {Ryan}, {Sachdev}, {Sadecki}, {Sadeghian}, {Salconi}, {Saleem},
  {Salemi}, {Samajdar}, {Sammut}, {Sanchez}, {Sandberg}, {Sandeen}, {Sanders},
  {Sassolas}, {Sathyaprakash}, {Saulson}, {Sauter}, {Savage}, {Sawadsky},
  {Schale}, {Schilling}, {Schmidt}, {Schmidt}, {Schnabel}, {Schofield},
  {Sch{\"o}nbeck}, {Schreiber}, {Schuette}, {Schutz}, {Scott}, {Scott},
  {Sellers}, {Sentenac}, {Sequino}, {Sergeev}, {Serna}, {Setyawati}, {Sevigny},
  {Shaddock}, {Shah}, {Shahriar}, {Shaltev}, {Shao}, {Shapiro}, {Shawhan},
  {Sheperd}, {Shoemaker}, {Shoemaker}, {Siellez}, {Siemens}, {Sigg}, {Silva},
  {Simakov}, {Singer}, {Singh}, {Singh}, {Singhal}, {Sintes}, {Slagmolen},
  {Smith}, {Smith}, {Smith}, {Son}, {Sorazu}, {Sorrentino}, {Souradeep},
  {Srivastava}, {Staley}, {Steinke}, {Steinlechner}, {Steinlechner},
  {Steinmeyer}, {Stephens}, {Stone}, {Strain}, {Straniero}, {Stratta},
  {Strauss}, {Strigin}, {Sturani}, {Stuver}, {Summerscales}, {Sun}, {Sutton},
  {Swinkels}, {Szczepa{\'n}czyk}, {Tacca}, {Talukder}, {Tanner}, {T{\'a}pai},
  {Tarabrin}, {Taracchini}, {Taylor}, {Theeg}, {Thirugnanasambandam}, {Thomas},
  {Thomas}, {Thomas}, {Thorne}, {Thorne}, {Thrane}, {Tiwari}, {Tiwari},
  {Tokmakov}, {Tomlinson}, {Tonelli}, {Torres}, {Torrie}, {T{\"o}yr{\"a}},
  {Travasso}, {Traylor}, {Trifir{\`o}}, {Tringali}, {Trozzo}, {Tse}, {Turconi},
  {Tuyenbayev}, {Ugolini}, {Unnikrishnan}, {Urban}, {Usman}, {Vahlbruch},
  {Vajente}, {Valdes}, {van Bakel}, {van Beuzekom}, {van den Brand}, {Van Den
  Broeck}, {Vander-Hyde}, {van der Schaaf}, {van Heijningen}, {van Veggel},
  {Vardaro}, {Vass}, {Vas{\'u}th}, {Vaulin}, {Vecchio}, {Vedovato}, {Veitch},
  {Veitch}, {Venkateswara}, {Verkindt}, {Vetrano}, {Vicer{\'e}}, {Vinciguerra},
  {Vine}, {Vinet}, {Vitale}, {Vo}, {Vocca}, {Vorvick}, {Voss}, {Vousden},
  {Vyatchanin}, {Wade}, {Wade}, {Wade}, {Walker}, {Wallace}, {Walsh}, {Wang},
  {Wang}, {Wang}, {Wang}, {Wang}, {Ward}, {Warner}, {Was}, {Weaver}, {Wei},
  {Weinert}, {Weinstein}, {Weiss}, {Welborn}, {Wen}, {We{\ss}els}, {Westphal},
  {Wette}, {Whelan}, {White}, {Whiting}, {Williams}, {Williamson}, {Willis},
  {Willke}, {Wimmer}, {Winkler}, {Wipf}, {Wittel}, {Woan}, {Worden}, {Wright},
  {Wu}, {Yablon}, {Yam}, {Yamamoto}, {Yancey}, {Yap}, {Yu}, {Yvert},
  {Zadro{\.z}ny}, {Zangrando}, {Zanolin}, {Zendri}, {Zevin}, {Zhang}, {Zhang},
  {Zhang}, {Zhang}, {Zhao}, {Zhou}, {Zhou}, {Zhu}, {Zucker}, {Zuraw},
  {Zweizig}, {LIGO Scientific Collaboration}, {Virgo Collaboration}, {Allison},
  {Bannister}, {Bell}, {Chatterjee}, {Chippendale}, {Edwards}, {Harvey-Smith},
  {Heywood}, {Hotan}, {Indermuehle}, {Marvil}, {McConnell}, {Murphy},
  {Popping}, {Reynolds}, {Sault}, {Voronkov}, {Whiting}, {Australian Square
  Kilometer Array Pathfinder (ASKAP Collaboration)}, {Castro-Tirado},
  {Cunniffe}, {Jel{\'\i}nek}, {Tello}, {Oates}, {Hu}, {Kub{\'a}nek}, {Guziy},
  {Castell{\'o}n}, {Garc{\'\i}a-Cerezo}, {Mu{\~n}oz}, {P{\'e}rez del Pulgar},
  {Castillo-Carri{\'o}n}, {Castro Cer{\'o}n}, {Hudec}, {Caballero-Garc{\'\i}a},
  {P{\'a}ta}, {Vitek}, {Adame}, {Konig}, {Rend{\'o}n}, {Mateo Sanguino},
  {Fern{\'a}ndez-Mu{\~n}oz}, {Yock}, {Rattenbury}, {Allen}, {Querel}, {Jeong},
  {Park}, {Bai}, {Cui}, {Fan}, {Wang}, {Hiriart}, {Lee}, {Claret},
  {S{\'a}nchez-Ram{\'\i}rez}, {Pandey}, {Mediavilla}, {Sabau-Graziati}, {BOOTES
  Collaboration}, {Abbott}, {Abdalla}, {Allam}, {Annis}, {Armstrong},
  {Benoit-L{\'e}vy}, {Berger}, {Bernstein}, {Bertin}, {Brout}, {Buckley-Geer},
  {Burke}, {Capozzi}, {Carretero}, {Castander}, {Chornock}, {Cowperthwaite},
  {Crocce}, {Cunha}, {D'Andrea}, {da Costa}, {Desai}, {Diehl}, {Dietrich},
  {Doctor}, {Drlica-Wagner}, {Drout}, {Eifler}, {Estrada}, {Evrard},
  {Fernandez}, {Finley}, {Flaugher}, {Foley}, {Fong}, {Fosalba}, {Fox},
  {Frieman}, {Fryer}, {Gaztanaga}, {Gerdes}, {Goldstein}, {Gruen}, {Gruendl},
  {Gutierrez}, {Herner}, {Honscheid}, {James}, {Johnson}, {Johnson},
  {Karliner}, {Kasen}, {Kent}, {Kessler}, {Kim}, {Kind}, {Kuehn}, {Kuropatkin},
  {Lahav}, {Li}, {Lima}, {Lin}, {Maia}, {Margutti}, {Marriner}, {Martini},
  {Matheson}, {Melchior}, {Metzger}, {Miller}, {Miquel}, {Neilsen}, {Nichol},
  {Nord}, {Nugent}, {Ogando}, {Petravick}, {Plazas}, {Quataert}, {Roe},
  {Romer}, {Roodman}, {Rosell}, {Rykoff}, {Sako}, {Sanchez}, {Scarpine},
  {Schindler}, {Schubnell}, {Scolnic}, {Sevilla-Noarbe}, {Sheldon}, {Smith},
  {Smith}, {Soares-Santos}, {Sobreira}, {Stebbins}, {Suchyta}, {Swanson},
  {Tarle}, {Thaler}, {Thomas}, {Thomas}, {Tucker}, {Vikram}, {Walker},
  {Wechsler}, {Wester}, {Yanny}, {Zhang}, {Zuntz}, {Dark Energy Survey
  Collaboration}, {Dark Energy Camera GW-EM Collaboration}, {Connaughton},
  {Burns}, {Goldstein}, {Briggs}, {Zhang}, {Hui}, {Jenke}, {Wilson-Hodge},
  {Bhat}, {Bissaldi}, {Cleveland}, {Fitzpatrick}, {Giles}, {Gibby}, {Greiner},
  {von Kienlin}, {Kippen}, {McBreen}, {Mailyan}, {Meegan}, {Paciesas},
  {Preece}, {Roberts}, {Sparke}, {Stanbro}, {Toelge}, {Veres}, {Yu},
  {Blackburn}, {Fermi GBM Collaboration}, {Ackermann}, {Ajello}, {Albert},
  {Anderson}, {Atwood}, {Axelsson}, {Baldini}, {Barbiellini}, {Bastieri},
  {Bellazzini}, {Bissaldi}, {Blandford}, {Bloom}, {Bonino}, {Bottacini},
  {Brandt}, {Bruel}, {Buson}, {Caliandro}, {Cameron}, {Caragiulo}, {Caraveo},
  {Cavazzuti}, {Charles}, {Chekhtman}, {Chiang}, {Chiaro}, {Ciprini},
  {Cohen-Tanugi}, {Cominsky}, {Costanza}, {Cuoco}, {D'Ammando}, {de Palma},
  {Desiante}, {Digel}, {Di Lalla}, {Di Mauro}, {Di Venere}, {Dom{\'\i}nguez},
  {Drell}, {Dubois}, {Favuzzi}, {Ferrara}, {Franckowiak}, {Fukazawa}, {Funk},
  {Fusco}, {Gargano}, {Gasparrini}, {Giglietto}, {Giommi}, {Giordano},
  {Giroletti}, {Glanzman}, {Godfrey}, {Gomez-Vargas}, {Green}, {Grenier},
  {Grove}, {Guiriec}, {Hadasch}, {Harding}, {Hays}, {Hewitt}, {Hill}, {Horan},
  {Jogler}, {J{\'o}hannesson}, {Johnson}, {Kensei}, {Kocevski}, {Kuss}, {La
  Mura}, {Larsson}, {Latronico}, {Li}, {Li}, {Longo}, {Loparco}, {Lovellette},
  {Lubrano}, {Magill}, {Maldera}, {Manfreda}, {Marelli}, {Mayer}, {Mazziotta},
  {McEnery}, {Meyer}, {Michelson}, {Mirabal}, {Mizuno}, {Moiseev}, {Monzani},
  {Moretti}, {Morselli}, {Moskalenko}, {Negro}, {Nuss}, {Ohsugi}, {Omodei},
  {Orienti}, {Orlando}, {Ormes}, {Paneque}, {Perkins}, {Pesce-Rollins},
  {Piron}, {Pivato}, {Porter}, {Racusin}, {Rain{\`o}}, {Rando}, {Razzaque},
  {Reimer}, {Reimer}, {Salvetti}, {Saz Parkinson}, {Sgr{\`o}}, {Simone},
  {Siskind}, {Spada}, {Spandre}, {Spinelli}, {Suson}, {Tajima}, {Thayer},
  {Thompson}, {Tibaldo}, {Torres}, {Troja}, {Uchiyama}, {Venters}, {Vianello},
  {Wood}, {Wood}, {Zhu}, {Zimmer}, {Fermi LAT Collaboration}, {Brocato},
  {Cappellaro}, {Covino}, {Grado}, {Nicastro}, {Palazzi}, {Pian}, {Amati},
  {Antonelli}, {Capaccioli}, {D'Avanzo}, {D'Elia}, {Getman}, {Giuffrida},
  {Iannicola}, {Limatola}, {Lisi}, {Marinoni}, {Marrese}, {Melandri},
  {Piranomonte}, {Possenti}, {Pulone}, {Rossi}, {Stamerra}, {Stella}, {Testa},
  {Tomasella}, {Yang}, {GRAvitational Wave Inaf TeAm (GRAWITA)}, {Bazzano},
  {Bozzo}, {Brandt}, {Courvoisier}, {Ferrigno}, {Hanlon}, {Kuulkers},
  {Laurent}, {Mereghetti}, {Roques}, {Savchenko}, {Ubertini}, {INTEGRAL
  Collaboration}, {Kasliwal}, {Singer}, {Cao}, {Duggan}, {Kulkarni},
  {Bhalerao}, {Miller}, {Barlow}, {Bellm}, {Manulis}, {Rana}, {Laher}, {Masci},
  {Surace}, {Rebbapragada}, {Cook}, {Van Sistine}, {Sesar}, {Perley},
  {Ferreti}, {Prince}, {Kendrick}, {Horesh}, {Intermediate Palomar Transient
  Factory (iPTF Collaboration)}, {Hurley}, {Golenetskii}, {Aptekar},
  {Frederiks}, {Svinkin}, {Rau}, {von Kienlin}, {Zhang}, {Smith}, {Cline},
  {Krimm}, {InterPlanetary Network}, {Abe}, {Doi}, {Fujisawa}, {Kawabata},
  {Morokuma}, {Motohara}, {Tanaka}, {Ohta}, {Yanagisawa}, {Yoshida}, {J-GEM
  Collaboration}, {Baltay}, {Rabinowitz}, {Ellman}, {Rostami}, {La Silla-QUEST
  Survey}, {Bersier}, {Bode}, {Collins}, {Copperwheat}, {Darnley}, {Galloway},
  {Gomboc}, {Kobayashi}, {Mazzali}, {Mundell}, {Piascik}, {Pollacco}, {Steele},
  {Ulaczyk}, {Liverpool Telescope Collaboration}, {Broderick}, {Fender},
  {Jonker}, {Rowlinson}, {Stappers}, {Wijers}, {Low Frequency Array (LOFAR
  Collaboration)}, {Lipunov}, {Gorbovskoy}, {Tyurina}, {Kornilov}, {Balanutsa},
  {Kuznetsov}, {Buckley}, {Rebolo}, {Serra-Ricart}, {Israelian}, {Budnev},
  {Gress}, {Ivanov}, {Poleshuk}, {Tlatov}, {Yurkov}, {MASTER Collaboration},
  {Kawai}, {Serino}, {Negoro}, {Nakahira}, {Mihara}, {Tomida}, {Ueno},
  {Tsunemi}, {Matsuoka}, {MAXI Collaboration}, {Croft}, {Feng}, {Franzen},
  {Gaensler}, {Johnston-Hollitt}, {Kaplan}, {Morales}, {Tingay}, {Wayth},
  {Williams}, {Murchison Wide-field Array (MWA Collaboration)}, {Smartt},
  {Chambers}, {Smith}, {Huber}, {Young}, {Wright}, {Schultz}, {Denneau},
  {Flewelling}, {Magnier}, {Primak}, {Rest}, {Sherstyuk}, {Stalder}, {Stubbs},
  {Tonry}, {Waters}, {Willman}, {Pan-STARRS Collaboration}, {Olivares E.},
  {Campbell}, {Kotak}, {Sollerman}, {Smith}, {Dennefeld}, {Anderson},
  {Botticella}, {Chen}, {Della Valle}, {Elias-Rosa}, {Fraser}, {Inserra},
  {Kankare}, {Kupfer}, {Harmanen}, {Galbany}, {Le Guillou}, {Lyman}, {Maguire},
  {Mitra}, {Nicholl}, {Razza}, {Terreran}, {Valenti}, {Gal-Yam}, {PESSTO
  Collaboration}, {{\'C}wiek}, {{\'C}wiok}, {Mankiewicz}, {Opiela}, {Zaremba},
  {{\.Z}arnecki}, {Pi of Sky Collaboration}, {Onken}, {Scalzo}, {Schmidt},
  {Wolf}, {Yuan}, {SkyMapper Collaboration}, {Evans}, {Kennea}, {Burrows},
  {Campana}, {Cenko}, {Giommi}, {Marshall}, {Nousek}, {O'Brien}, {Osborne},
  {Palmer}, {Perri}, {Siegel}, {Tagliaferri}, {Swift Collaboration}, {Klotz},
  {Turpin}, {Laugier}, {TAROT}, Collaboration, {Beroiz}, {Pe{\~n}uela},
  {Macri}, {Oelkers}, {Lambas}, {Vrech}, {Cabral}, {Colazo}, {Dominguez},
  {Sanchez}, {Gurovich}, {Lares}, {Marshall}, {DePoy}, {Padilla}, {Pereyra},
  {Benacquista}, {TOROS Collaboration}, {Tanvir}, {Wiersema}, {Levan},
  {Steeghs}, {Hjorth}, {Fynbo}, {Malesani}, {Milvang-Jensen}, {Watson},
  {Irwin}, {Fernandez}, {McMahon}, {Banerji}, {Gonzalez-Solares}, {Schulze},
  {de Ugarte Postigo}, {Thoene}, {Cano}, {Rosswog}, \& {VISTA
  Collaboration}}]{2016ApJ...826L..13A}
---. 2016{\natexlab{b}}, \apjl, 826, L13, \dodoi{10.3847/2041-8205/826/1/L13}

\bibitem[{{Abbott} {et~al.}(2017){Abbott}, {Abbott}, {Abbott}, {Acernese},
  {Ackley}, {Adams}, {Adams}, {Addesso}, {Adhikari}, {Adya}, {Affeldt},
  {Afrough}, {Agarwal}, {Agathos}, {Agatsuma}, {Aggarwal}, {Aguiar}, {Aiello},
  {Ain}, {Ajith}, {Allen}, {Allen}, {Allocca}, {Altin}, {Amato}, {Ananyeva},
  {Anderson}, {Anderson}, {Angelova}, {Antier}, {Appert}, {Arai}, {Araya},
  {Areeda}, {Arnaud}, {Arun}, {Ascenzi}, {Ashton}, {Ast}, {Aston}, {Astone},
  {Atallah}, {Aufmuth}, {Aulbert}, {AultONeal}, {Austin}, {Avila-Alvarez},
  {Babak}, {Bacon}, {Bader}, {Bae}, {Bailes}, {Baker}, {Baldaccini},
  {Ballardin}, {Ballmer}, {Banagiri}, {Barayoga}, {Barclay}, {Barish},
  {Barker}, {Barkett}, {Barone}, {Barr}, {Barsotti}, {Barsuglia}, {Barta},
  {Barthelmy}, {Bartlett}, {Bartos}, {Bassiri}, {Basti}, {Batch}, {Bawaj},
  {Bayley}, {Bazzan}, {B{\'e}csy}, {Beer}, {Bejger}, {Belahcene}, {Bell},
  {Berger}, {Bergmann}, {Bernuzzi}, {Bero}, {Berry}, {Bersanetti}, {Bertolini},
  {Betzwieser}, {Bhagwat}, {Bhandare}, {Bilenko}, {Billingsley}, {Billman},
  {Birch}, {Birney}, {Birnholtz}, {Biscans}, {Biscoveanu}, {Bisht}, {Bitossi},
  {Biwer}, {Bizouard}, {Blackburn}, {Blackman}, {Blair}, {Blair}, {Blair},
  {Bloemen}, {Bock}, {Bode}, {Boer}, {Bogaert}, {Bohe}, {Bondu}, {Bonilla},
  {Bonnand}, {Boom}, {Bork}, {Boschi}, {Bose}, {Bossie}, {Bouffanais}, {Bozzi},
  {Bradaschia}, {Brady}, {Branchesi}, {Brau}, {Briant}, {Brillet}, {Brinkmann},
  {Brisson}, {Brockill}, {Broida}, {Brooks}, {Brown}, {Brown}, {Brunett},
  {Buchanan}, {Buikema}, {Bulik}, {Bulten}, {Buonanno}, {Buskulic}, {Buy},
  {Byer}, {Cabero}, {Cadonati}, {Cagnoli}, {Cahillane}, {Calder{\'o}n
  Bustillo}, {Callister}, {Calloni}, {Camp}, {Canepa}, {Canizares}, {Cannon},
  {Cao}, {Cao}, {Capano}, {Capocasa}, {Carbognani}, {Caride}, {Carney},
  {Carullo}, {Casanueva Diaz}, {Casentini}, {Caudill}, {Cavagli{\`a}},
  {Cavalier}, {Cavalieri}, {Cella}, {Cepeda}, {Cerd{\'a}-Dur{\'a}n},
  {Cerretani}, {Cesarini}, {Chamberlin}, {Chan}, {Chao}, {Charlton}, {Chase},
  {Chassande-Mottin}, {Chatterjee}, {Chatziioannou}, {Cheeseboro}, {Chen},
  {Chen}, {Chen}, {Cheng}, {Chia}, {Chincarini}, {Chiummo}, {Chmiel}, {Cho},
  {Cho}, {Chow}, {Christensen}, {Chu}, {Chua}, {Chua}, {Chung}, {Chung},
  {Ciani}, {Ciolfi}, {Cirelli}, {Cirone}, {Clara}, {Clark}, {Clearwater},
  {Cleva}, {Cocchieri}, {Coccia}, {Cohadon}, {Cohen}, {Colla}, {Collette},
  {Cominsky}, {Constancio}, {Conti}, {Cooper}, {Corban}, {Corbitt},
  {Cordero-Carri{\'o}n}, {Corley}, {Cornish}, {Corsi}, {Cortese}, {Costa},
  {Coughlin}, {Coughlin}, {Coulon}, {Countryman}, {Couvares}, {Covas}, {Cowan},
  {Coward}, {Cowart}, {Coyne}, {Coyne}, {Creighton}, {Creighton}, {Cripe},
  {Crowder}, {Cullen}, {Cumming}, {Cunningham}, {Cuoco}, {Dal Canton},
  {D{\'a}lya}, {Danilishin}, {D'Antonio}, {Danzmann}, {Dasgupta}, {Da Silva
  Costa}, {Dattilo}, {Dave}, {Davier}, {Davis}, {Daw}, {Day}, {De}, {DeBra},
  {Degallaix}, {De Laurentis}, {Del{\'e}glise}, {Del Pozzo}, {Demos}, {Denker},
  {Dent}, {De Pietri}, {Dergachev}, {De Rosa}, {DeRosa}, {De Rossi}, {DeSalvo},
  {de Varona}, {Devenson}, {Dhurandhar}, {D{\'\i}az}, {Dietrich}, {Di Fiore},
  {Di Giovanni}, {Di Girolamo}, {Di Lieto}, {Di Pace}, {Di Palma}, {Di Renzo},
  {Doctor}, {Dolique}, {Donovan}, {Dooley}, {Doravari}, {Dorrington},
  {Douglas}, {Dovale {\'A}lvarez}, {Downes}, {Drago}, {Dreissigacker},
  {Driggers}, {Du}, {Ducrot}, {Dudi}, {Dupej}, {Dwyer}, {Edo}, {Edwards},
  {Effler}, {Eggenstein}, {Ehrens}, {Eichholz}, {Eikenberry}, {Eisenstein},
  {Essick}, {Estevez}, {Etienne}, {Etzel}, {Evans}, {Evans}, {Factourovich},
  {Fafone}, {Fair}, {Fairhurst}, {Fan}, {Farinon}, {Farr}, {Farr},
  {Fauchon-Jones}, {Favata}, {Fays}, {Fee}, {Fehrmann}, {Feicht}, {Fejer},
  {Fernandez-Galiana}, {Ferrante}, {Ferreira}, {Ferrini}, {Fidecaro},
  {Finstad}, {Fiori}, {Fiorucci}, {Fishbach}, {Fisher}, {Fitz-Axen},
  {Flaminio}, {Fletcher}, {Fong}, {Font}, {Forsyth}, {Forsyth}, {Fournier},
  {Frasca}, {Frasconi}, {Frei}, {Freise}, {Frey}, {Frey}, {Fries}, {Fritschel},
  {Frolov}, {Fulda}, {Fyffe}, {Gabbard}, {Gadre}, {Gaebel}, {Gair},
  {Gammaitoni}, {Ganija}, {Gaonkar}, {Garcia-Quiros}, {Garufi}, {Gateley},
  {Gaudio}, {Gaur}, {Gayathri}, {Gehrels}, {Gemme}, {Genin}, {Gennai},
  {George}, {George}, {Gergely}, {Germain}, {Ghonge}, {Ghosh}, {Ghosh},
  {Ghosh}, {Giaime}, {Giardina}, {Giazotto}, {Gill}, {Glover}, {Goetz},
  {Goetz}, {Gomes}, {Goncharov}, {Gonz{\'a}lez}, {Gonzalez Castro},
  {Gopakumar}, {Gorodetsky}, {Gossan}, {Gosselin}, {Gouaty}, {Grado}, {Graef},
  {Granata}, {Grant}, {Gras}, {Gray}, {Greco}, {Green}, {Gretarsson}, {Groot},
  {Grote}, {Grunewald}, {Gruning}, {Guidi}, {Guo}, {Gupta}, {Gupta}, {Gushwa},
  {Gustafson}, {Gustafson}, {Halim}, {Hall}, {Hall}, {Hamilton}, {Hammond},
  {Haney}, {Hanke}, {Hanks}, {Hanna}, {Hannam}, {Hannuksela}, {Hanson},
  {Hardwick}, {Harms}, {Harry}, {Harry}, {Hart}, {Haster}, {Haughian}, {Healy},
  {Heidmann}, {Heintze}, {Heitmann}, {Hello}, {Hemming}, {Hendry}, {Heng},
  {Hennig}, {Heptonstall}, {Heurs}, {Hild}, {Hinderer}, {Ho}, {Hoak}, {Hofman},
  {Holt}, {Holz}, {Hopkins}, {Horst}, {Hough}, {Houston}, {Howell}, {Hreibi},
  {Hu}, {Huerta}, {Huet}, {Hughey}, {Husa}, {Huttner}, {Huynh-Dinh}, {Indik},
  {Inta}, {Intini}, {Isa}, {Isac}, {Isi}, {Iyer}, {Izumi}, {Jacqmin}, {Jani},
  {Jaranowski}, {Jawahar}, {Jim{\'e}nez-Forteza}, {Johnson},
  {Johnson-McDaniel}, {Jones}, {Jones}, {Jonker}, {Ju}, {Junker}, {Kalaghatgi},
  {Kalogera}, {Kamai}, {Kandhasamy}, {Kang}, {Kanner}, {Kapadia}, {Karki},
  {Karvinen}, {Kasprzack}, {Kastaun}, {Katolik}, {Katsavounidis}, {Katzman},
  {Kaufer}, {Kawabe}, {K{\'e}f{\'e}lian}, {Keitel}, {Kemball}, {Kennedy},
  {Kent}, {Key}, {Khalili}, {Khan}, {Khan}, {Khan}, {Khazanov}, {Kijbunchoo},
  {Kim}, {Kim}, {Kim}, {Kim}, {Kim}, {Kim}, {Kimbrell}, {King}, {King},
  {Kinley-Hanlon}, {Kirchhoff}, {Kissel}, {Kleybolte}, {Klimenko}, {Knowles},
  {Koch}, {Koehlenbeck}, {Koley}, {Kondrashov}, {Kontos}, {Korobko}, {Korth},
  {Kowalska}, {Kozak}, {Kr{\"a}mer}, {Kringel}, {Krishnan}, {Kr{\'o}lak},
  {Kuehn}, {Kumar}, {Kumar}, {Kumar}, {Kuo}, {Kutynia}, {Kwang}, {Lackey},
  {Lai}, {Landry}, {Lang}, {Lange}, {Lantz}, {Lanza}, {Larson},
  {Lartaux-Vollard}, {Lasky}, {Laxen}, {Lazzarini}, {Lazzaro}, {Leaci},
  {Leavey}, {Lee}, {Lee}, {Lee}, {Lee}, {Lee}, {Lehmann}, {Lenon}, {Leon},
  {Leonardi}, {Leroy}, {Letendre}, {Levin}, {Li}, {Linker}, {Littenberg},
  {Liu}, {Liu}, {Lo}, {Lockerbie}, {London}, {Lord}, {Lorenzini}, {Loriette},
  {Lormand}, {Losurdo}, {Lough}, {Lousto}, {Lovelace}, {L{\"u}ck}, {Lumaca},
  {Lundgren}, {Lynch}, {Ma}, {Macas}, {Macfoy}, {Machenschalk}, {MacInnis},
  {Macleod}, {Maga{\~n}a Hernandez}, {Maga{\~n}a-Sandoval}, {Maga{\~n}a
  Zertuche}, {Magee}, {Majorana}, {Maksimovic}, {Man}, {Mandic}, {Mangano},
  {Mansell}, {Manske}, {Mantovani}, {Marchesoni}, {Marion}, {M{\'a}rka},
  {M{\'a}rka}, {Markakis}, {Markosyan}, {Markowitz}, {Maros}, {Marquina},
  {Marsh}, {Martelli}, {Martellini}, {Martin}, {Martin}, {Martynov}, {Marx},
  {Mason}, {Massera}, {Masserot}, {Massinger}, {Masso-Reid}, {Mastrogiovanni},
  {Matas}, {Matichard}, {Matone}, {Mavalvala}, {Mazumder}, {McCarthy},
  {McClelland}, {McCormick}, {McCuller}, {McGuire}, {McIntyre}, {McIver},
  {McManus}, {McNeill}, {McRae}, {McWilliams}, {Meacher}, {Meadors}, {Mehmet},
  {Meidam}, {Mejuto-Villa}, {Melatos}, {Mendell}, {Mercer}, {Merilh},
  {Merzougui}, {Meshkov}, {Messenger}, {Messick}, {Metzdorff}, {Meyers},
  {Miao}, {Michel}, {Middleton}, {Mikhailov}, {Milano}, {Miller}, {Miller},
  {Miller}, {Millhouse}, {Milovich-Goff}, {Minazzoli}, {Minenkov}, {Ming},
  {Mishra}, {Mitra}, {Mitrofanov}, {Mitselmakher}, {Mittleman}, {Moffa},
  {Moggi}, {Mogushi}, {Mohan}, {Mohapatra}, {Molina}, {Montani}, {Moore},
  {Moraru}, {Moreno}, {Morisaki}, {Morriss}, {Mours}, {Mow-Lowry}, {Mueller},
  {Muir}, {Mukherjee}, {Mukherjee}, {Mukherjee}, {Mukund}, {Mullavey}, {Munch},
  {Mu{\~n}iz}, {Muratore}, {Murray}, {Nagar}, {Napier}, {Nardecchia},
  {Naticchioni}, {Nayak}, {Neilson}, {Nelemans}, {Nelson}, {Nery}, {Neunzert},
  {Nevin}, {Newport}, {Newton}, {Ng}, {Nguyen}, {Nguyen}, {Nichols}, {Nielsen},
  {Nissanke}, {Nitz}, {Noack}, {Nocera}, {Nolting}, {North}, {Nuttall},
  {Oberling}, {O'Dea}, {Ogin}, {Oh}, {Oh}, {Ohme}, {Okada}, {Oliver},
  {Oppermann}, {Oram}, {O'Reilly}, {Ormiston}, {Ortega}, {O'Shaughnessy},
  {Ossokine}, {Ottaway}, {Overmier}, {Owen}, {Pace}, {Page}, {Page}, {Pai},
  {Pai}, {Palamos}, {Palashov}, {Palomba}, {Pal-Singh}, {Pan}, {Pan}, {Pang},
  {Pang}, {Pankow}, {Pannarale}, {Pant}, {Paoletti}, {Paoli}, {Papa}, {Parida},
  {Parker}, {Pascucci}, {Pasqualetti}, {Passaquieti}, {Passuello}, {Patil},
  {Patricelli}, {Pearlstone}, {Pedraza}, {Pedurand}, {Pekowsky}, {Pele},
  {Penn}, {Perez}, {Perreca}, {Perri}, {Pfeiffer}, {Phelps}, {Piccinni},
  {Pichot}, {Piergiovanni}, {Pierro}, {Pillant}, {Pinard}, {Pinto}, {Pirello},
  {Pitkin}, {Poe}, {Poggiani}, {Popolizio}, {Porter}, {Post}, {Powell},
  {Prasad}, {Pratt}, {Pratten}, {Predoi}, {Prestegard}, {Prijatelj},
  {Principe}, {Privitera}, {Prix}, {Prodi}, {Prokhorov}, {Puncken}, {Punturo},
  {Puppo}, {P{\"u}rrer}, {Qi}, {Quetschke}, {Quintero}, {Quitzow-James},
  {Raab}, {Rabeling}, {Radkins}, {Raffai}, {Raja}, {Rajan}, {Rajbhandari},
  {Rakhmanov}, {Ramirez}, {Ramos-Buades}, {Rapagnani}, {Raymond}, {Razzano},
  {Read}, {Regimbau}, {Rei}, {Reid}, {Reitze}, {Ren}, {Reyes}, {Ricci},
  {Ricker}, {Rieger}, {Riles}, {Rizzo}, {Robertson}, {Robie}, {Robinet},
  {Rocchi}, {Rolland}, {Rollins}, {Roma}, {Romano}, {Romano}, {Romel}, {Romie},
  {Rosi{\'n}ska}, {Ross}, {Rowan}, {R{\"u}diger}, {Ruggi}, {Rutins}, {Ryan},
  {Sachdev}, {Sadecki}, {Sadeghian}, {Sakellariadou}, {Salconi}, {Saleem},
  {Salemi}, {Samajdar}, {Sammut}, {Sampson}, {Sanchez}, {Sanchez},
  {Sanchis-Gual}, {Sandberg}, {Sanders}, {Sassolas}, {Sathyaprakash},
  {Saulson}, {Sauter}, {Savage}, {Sawadsky}, {Schale}, {Scheel}, {Scheuer},
  {Schmidt}, {Schmidt}, {Schnabel}, {Schofield}, {Sch{\"o}nbeck}, {Schreiber},
  {Schuette}, {Schulte}, {Schutz}, {Schwalbe}, {Scott}, {Scott}, {Seidel},
  {Sellers}, {Sengupta}, {Sentenac}, {Sequino}, {Sergeev}, {Shaddock},
  {Shaffer}, {Shah}, {Shahriar}, {Shaner}, {Shao}, {Shapiro}, {Shawhan},
  {Sheperd}, {Shoemaker}, {Shoemaker}, {Siellez}, {Siemens}, {Sieniawska},
  {Sigg}, {Silva}, {Singer}, {Singh}, {Singhal}, {Sintes}, {Slagmolen},
  {Smith}, {Smith}, {Smith}, {Somala}, {Son}, {Sonnenberg}, {Sorazu},
  {Sorrentino}, {Souradeep}, {Spencer}, {Srivastava}, {Staats}, {Staley},
  {Steinke}, {Steinlechner}, {Steinlechner}, {Steinmeyer}, {Stevenson},
  {Stone}, {Stops}, {Strain}, {Stratta}, {Strigin}, {Strunk}, {Sturani},
  {Stuver}, {Summerscales}, {Sun}, {Sunil}, {Suresh}, {Sutton}, {Swinkels},
  {Szczepa{\'n}czyk}, {Tacca}, {Tait}, {Talbot}, {Talukder}, {Tanner},
  {T{\'a}pai}, {Taracchini}, {Tasson}, {Taylor}, {Taylor}, {Tewari}, {Theeg},
  {Thies}, {Thomas}, {Thomas}, {Thomas}, {Thorne}, {Thorne}, {Thrane},
  {Tiwari}, {Tiwari}, {Tokmakov}, {Toland}, {Tonelli}, {Tornasi},
  {Torres-Forn{\'e}}, {Torrie}, {T{\"o}yr{\"a}}, {Travasso}, {Traylor},
  {Trinastic}, {Tringali}, {Trozzo}, {Tsang}, {Tse}, {Tso}, {Tsukada}, {Tsuna},
  {Tuyenbayev}, {Ueno}, {Ugolini}, {Unnikrishnan}, {Urban}, {Usman},
  {Vahlbruch}, {Vajente}, {Valdes}, {Vallisneri}, {van Bakel}, {van Beuzekom},
  {van den Brand}, {Van Den Broeck}, {Vander-Hyde}, {van der Schaaf}, {van
  Heijningen}, {van Veggel}, {Vardaro}, {Varma}, {Vass}, {Vas{\'u}th},
  {Vecchio}, {Vedovato}, {Veitch}, {Veitch}, {Venkateswara}, {Venugopalan},
  {Verkindt}, {Vetrano}, {Vicer{\'e}}, {Viets}, {Vinciguerra}, {Vine}, {Vinet},
  {Vitale}, {Vo}, {Vocca}, {Vorvick}, {Vyatchanin}, {Wade}, {Wade}, {Wade},
  {Walet}, {Walker}, {Wallace}, {Walsh}, {Wang}, {Wang}, {Wang}, {Wang},
  {Wang}, {Ward}, {Warner}, {Was}, {Watchi}, {Weaver}, {Wei}, {Weinert},
  {Weinstein}, {Weiss}, {Wen}, {Wessel}, {We{\ss}els}, {Westerweck},
  {Westphal}, {Wette}, {Whelan}, {Whitcomb}, {Whiting}, {Whittle}, {Wilken},
  {Williams}, {Williams}, {Williamson}, {Willis}, {Willke}, {Wimmer},
  {Winkler}, {Wipf}, {Wittel}, {Woan}, {Woehler}, {Wofford}, {Wong}, {Worden},
  {Wright}, {Wu}, {Wysocki}, {Xiao}, {Yamamoto}, {Yancey}, {Yang}, {Yap},
  {Yazback}, {Yu}, {Yu}, {Yvert}, {Zadro{\.Z}ny}, {Zanolin}, {Zelenova},
  {Zendri}, {Zevin}, {Zhang}, {Zhang}, {Zhang}, {Zhang}, {Zhao}, {Zhou},
  {Zhou}, {Zhu}, {Zhu}, {Zimmerman}, {Zucker}, {Zweizig}, {LIGO Scientific
  Collaboration}, \& {Virgo Collaboration}}]{2017PhRvL.119p1101A}
---. 2017, \prl, 119, 161101, \dodoi{10.1103/PhysRevLett.119.161101}

\bibitem[{{Abbott} {et~al.}(2020){Abbott}, {Abbott}, {Abbott}, {Abraham},
  {Acernese}, {Ackley}, {Adams}, {Adya}, {Affeldt}, {Agathos}, {Agatsuma},
  {Aggarwal}, {Aguiar}, {Aiello}, {Ain}, {Ajith}, {Akutsu}, {Allen}, {Allocca},
  {Aloy}, {Altin}, {Amato}, {Ananyeva}, {Anderson}, {Anderson}, {Ando},
  {Angelova}, {Antier}, {Appert}, {Arai}, {Arai}, {Arai}, {Araki}, {Araya},
  {Araya}, {Areeda}, {Ar{\`e}ne}, {Aritomi}, {Arnaud}, {Arun}, {Ascenzi},
  {Ashton}, {Aso}, {Aston}, {Astone}, {Aubin}, {Aufmuth}, {Aultoneal},
  {Austin}, {Avendano}, {Avila-Alvarez}, {Babak}, {Bacon}, {Badaracco},
  {Bader}, {Bae}, {Bae}, {Baiotti}, {Bajpai}, {Baker}, {Baldaccini},
  {Ballardin}, {Ballmer}, {Banagiri}, {Barayoga}, {Barclay}, {Barish},
  {Barker}, {Barkett}, {Barnum}, {Barone}, {Barr}, {Barsotti}, {Barsuglia},
  {Barta}, {Bartlett}, {Barton}, {Bartos}, {Bassiri}, {Basti}, {Bawaj},
  {Bayley}, {Bazzan}, {B{\'e}csy}, {Bejger}, {Belahcene}, {Bell}, {Beniwal},
  {Berger}, {Bergmann}, {Bernuzzi}, {Bero}, {Berry}, {Bersanetti}, {Bertolini},
  {Betzwieser}, {Bhandare}, {Bidler}, {Bilenko}, {Bilgili}, {Billingsley},
  {Birch}, {Birney}, {Birnholtz}, {Biscans}, {Biscoveanu}, {Bisht}, {Bitossi},
  {Bizouard}, {Blackburn}, {Blair}, {Blair}, {Blair}, {Bloemen}, {Bode},
  {Boer}, {Boetzel}, {Bogaert}, {Bondu}, {Bonilla}, {Bonnand}, {Booker},
  {Boom}, {Booth}, {Bork}, {Boschi}, {Bose}, {Bossie}, {Bossilkov}, {Bosveld},
  {Bouffanais}, {Bozzi}, {Bradaschia}, {Brady}, {Bramley}, {Branchesi}, {Brau},
  {Briant}, {Briggs}, {Brighenti}, {Brillet}, {Brinkmann}, {Brisson},
  {Brockill}, {Brooks}, {Brown}, {Brown}, {Brunett}, {Buikema}, {Bulik},
  {Bulten}, {Buonanno}, {Buskulic}, {Buy}, {Byer}, {Cabero}, {Cadonati},
  {Cagnoli}, {Cahillane}, {Bustillo}, {Callister}, {Calloni}, {Camp},
  {Campbell}, {Canepa}, {Cannon}, {Cannon}, {Cao}, {Cao}, {Capocasa},
  {Carbognani}, {Caride}, {Carney}, {Carullo}, {Casanueva Diaz}, {Casentini},
  {Caudill}, {Cavagli{\`a}}, {Cavalier}, {Cavalieri}, {Cella},
  {Cerd{\'a}-Dur{\'a}n}, {Cerretani}, {Cesarini}, {Chaibi}, {Chakravarti},
  {Chamberlin}, {Chan}, {Chan}, {Chao}, {Charlton}, {Chase},
  {Chassande-Mottin}, {Chatterjee}, {Chaturvedi}, {Chatziioannou},
  {Cheeseboro}, {Chen}, {Chen}, {Chen}, {Chen}, {Chen}, {Chen}, {Cheng},
  {Cheong}, {Chia}, {Chincarini}, {Chiummo}, {Cho}, {Cho}, {Cho},
  {Christensen}, {Chu}, {Chu}, {Chu}, {Chua}, {Chung}, {Chung}, {Ciani},
  {Ciobanu}, {Ciolfi}, {Cipriano}, {Cirone}, {Clara}, {Clark}, {Clearwater},
  {Cleva}, {Cocchieri}, {Coccia}, {Cohadon}, {Cohen}, {Colgan}, {Colleoni},
  {Collette}, {Collins}, {Cominsky}, {Constancio}, {Conti}, {Cooper}, {Corban},
  {Corbitt}, {Cordero-Carri{\'o}n}, {Corley}, {Cornish}, {Corsi}, {Cortese},
  {Costa}, {Cotesta}, {Coughlin}, {Coughlin}, {Coulon}, {Countryman},
  {Couvares}, {Covas}, {Cowan}, {Coward}, {Cowart}, {Coyne}, {Coyne},
  {Creighton}, {Creighton}, {Cripe}, {Croquette}, {Crowder}, {Cullen},
  {Cumming}, {Cunningham}, {Cuoco}, {Dal Canton}, {D{\'a}lya}, {Danilishin},
  {D'Antonio}, {Danzmann}, {Dasgupta}, {da Silva Costa}, {Datrier}, {Dattilo},
  {Dave}, {Davier}, {Davis}, {Daw}, {Debra}, {Deenadayalan}, {Degallaix}, {de
  Laurentis}, {Del{\'e}glise}, {Pozzo}, {Demarchi}, {Demos}, {Dent}, {de
  Pietri}, {Derby}, {De Rosa}, {de Rossi}, {Desalvo}, {de Varona},
  {Dhurandhar}, {D{\'\i}az}, {Dietrich}, {di Fiore}, {di Giovanni}, {di
  Girolamo}, {di Lieto}, {Ding}, {di Pace}, {di Palma}, {di Renzo}, {Dmitriev},
  {Doctor}, {Doi}, {Donovan}, {Dooley}, {Doravari}, {Dorrington}, {Downes},
  {Drago}, {Driggers}, {Du}, {Ducoin}, {Dupej}, {Dwyer}, {Easter}, {Edo},
  {Edwards}, {Effler}, {Eguchi}, {Ehrens}, {Eichholz}, {Eikenberry},
  {Eisenmann}, {Eisenstein}, {Enomoto}, {Essick}, {Estelles}, {Estevez},
  {Etienne}, {Etzel}, {Evans}, {Evans}, {Fafone}, {Fair}, {Fairhurst}, {Fan},
  {Farinon}, {Farr}, {Farr}, {Fauchon-Jones}, {Favata}, {Fays}, {Fazio}, {Fee},
  {Feicht}, {Fejer}, {Feng}, {Fernandez-Galiana}, {Ferrante}, {Ferreira},
  {Ferreira}, {Ferrini}, {Fidecaro}, {Fiori}, {Fiorucci}, {Fishbach}, {Fisher},
  {Fishner}, {Fitz-Axen}, {Flaminio}, {Fletcher}, {Flynn}, {Fong}, {Font},
  {Forsyth}, {Fournier}, {Frasca}, {Frasconi}, {Frei}, {Freise}, {Frey},
  {Frey}, {Fritschel}, {Frolov}, {Fujii}, {Fukunaga}, {Fukushima}, {Fulda},
  {Fyffe}, {Gabbard}, {Gadre}, {Gaebel}, {Gair}, {Gammaitoni}, {Ganija},
  {Gaonkar}, {Garcia}, {Garc{\'\i}a-Quir{\'o}s}, {Garufi}, {Gateley}, {Gaudio},
  {Gaur}, {Gayathri}, {Ge}, {Gemme}, {Genin}, {Gennai}, {George}, {George},
  {Gergely}, {Germain}, {Ghonge}, {Ghosh}, {Ghosh}, {Ghosh}, {Giacomazzo},
  {Giaime}, {Giardina}, {Giazotto}, {Gill}, {Giordano}, {Glover}, {Godwin},
  {Goetz}, {Goetz}, {Goncharov}, {Gonz{\'a}lez}, {Gonzalez Castro},
  {Gopakumar}, {Gorodetsky}, {Gossan}, {Gosselin}, {Gouaty}, {Grado}, {Graef},
  {Granata}, {Grant}, {Gras}, {Grassia}, {Gray}, {Gray}, {Greco}, {Green},
  {Green}, {Gretarsson}, {Groot}, {Grote}, {Grunewald}, {Gruning}, {Guidi},
  {Gulati}, {Guo}, {Gupta}, {Gupta}, {Gustafson}, {Gustafson}, {Haegel},
  {Hagiwara}, {Haino}, {Halim}, {Hall}, {Hall}, {Hamilton}, {Hammond}, {Haney},
  {Hanke}, {Hanks}, {Hanna}, {Hannam}, {Hannuksela}, {Hanson}, {Hardwick},
  {Haris}, {Harms}, {Harry}, {Harry}, {Hasegawa}, {Haster}, {Haughian},
  {Hayakawa}, {Hayama}, {Hayes}, {Healy}, {Heidmann}, {Heintze}, {Heitmann},
  {Hello}, {Hemming}, {Hendry}, {Heng}, {Hennig}, {Heptonstall}, {Heurs},
  {Hild}, {Himemoto}, {Hinderer}, {Hiranuma}, {Hirata}, {Hirose}, {Hoak},
  {Hochheim}, {Hofman}, {Holgado}, {Holland}, {Holt}, {Holz}, {Hong},
  {Hopkins}, {Horst}, {Hough}, {Howell}, {Hoy}, {Hreibi}, {Hsieh}, {Huang},
  {Huang}, {Huang}, {Huerta}, {Huet}, {Hughey}, {Hulko}, {Husa}, {Huttner},
  {Huynh-Dinh}, {Idzkowski}, {Iess}, {Ikenoue}, {Imam}, {Inayoshi}, {Ingram},
  {Inoue}, {Inta}, {Intini}, {Ioka}, {Irwin}, {Isa}, {Isac}, {Isi}, {Itoh},
  {Iyer}, {Izumi}, {Jacqmin}, {Jadhav}, {Jani}, {Janthalur}, {Jaranowski},
  {Jenkins}, {Jiang}, {Johnson}, {Jones}, {Jones}, {Jones}, {Jonker}, {Ju},
  {Jung}, {Jung}, {Junker}, {Kajita}, {Kalaghatgi}, {Kalogera}, {Kamai},
  {Kamiizumi}, {Kanda}, {Kandhasamy}, {Kang}, {Kanner}, {Kapadia}, {Karki},
  {Karvinen}, {Kashyap}, {Kasprzack}, {Katsanevas}, {Katsavounidis}, {Katzman},
  {Kaufer}, {Kawabe}, {Kawaguchi}, {Kawai}, {Kawasaki}, {Keerthana},
  {K{\'e}f{\'e}lian}, {Keitel}, {Kennedy}, {Key}, {Khalili}, {Khan}, {Khan},
  {Khan}, {Khan}, {Khazanov}, {Khursheed}, {Kijbunchoo}, {Kim}, {Kim}, {Kim},
  {Kim}, {Kim}, {Kim}, {Kim}, {Kim}, {Kimball}, {Kimura}, {King}, {King},
  {Kinley-Hanlon}, {Kirchhoff}, {Kissel}, {Kita}, {Kitazawa}, {Kleybolte},
  {Klika}, {Klimenko}, {Knowles}, {Knyazev}, {Koch}, {Koehlenbeck}, {Koekoek},
  {Kojima}, {Kokeyama}, {Koley}, {Komori}, {Kondrashov}, {Kong}, {Kontos},
  {Koper}, {Korobko}, {Korth}, {Kotake}, {Kowalska}, {Kozak}, {Kozakai},
  {Kozu}, {Kringel}, {Krishnendu}, {Kr{\'o}lak}, {Kuehn}, {Kumar}, {Kumar},
  {Kumar}, {Kumar}, {Kumar}, {Kume}, {Kuo}, {Kuo}, {Kuo}, {Kuroyanagi},
  {Kusayanagi}, {Kutynia}, {Kwak}, {Kwang}, {Lackey}, {Lai}, {Lam}, {Landry},
  {Lane}, {Lang}, {Lange}, {Lantz}, {Lanza}, {Lartaux-Vollard}, {Lasky},
  {Laxen}, {Lazzarini}, {Lazzaro}, {Leaci}, {Leavey}, {Lecoeuche}, {Lee},
  {Lee}, {Lee}, {Lee}, {Lee}, {Lee}, {Lee}, {Lehmann}, {Lenon}, {Leonardi},
  {Leroy}, {Letendre}, {Levin}, {Li}, {Li}, {Li}, {Li}, {Lin}, {Lin}, {Lin},
  {Lin}, {Linde}, {Linker}, {Littenberg}, {Liu}, {Liu}, {Liu}, {Lo},
  {Lockerbie}, {London}, {Longo}, {Lorenzini}, {Loriette}, {Lormand},
  {Losurdo}, {Lough}, {Lousto}, {Lovelace}, {Lower}, {L{\"u}ck}, {Lumaca},
  {Lundgren}, {Luo}, {Lynch}, {Ma}, {Macas}, {Macfoy}, {Macinnis}, {MacLeod},
  {Macquet}, {Maga{\~n}a-Sandoval}, {Zertuche}, {Magee}, {Majorana},
  {Maksimovic}, {Malik}, {Man}, {Mandic}, {Mangano}, {Mansell}, {Manske},
  {Mantovani}, {Marchesoni}, {Marchio}, {Marion}, {M{\'a}rka}, {M{\'a}rka},
  {Markakis}, {Markosyan}, {Markowitz}, {Maros}, {Marquina}, {Marsat},
  {Martelli}, {Martin}, {Martin}, {Martynov}, {Mason}, {Massera}, {Masserot},
  {Massinger}, {Masso-Reid}, {Mastrogiovanni}, {Matas}, {Matichard}, {Matone},
  {Mavalvala}, {Mazumder}, {McCann}, {McCarthy}, {McClelland}, {McCormick},
  {McCuller}, {McGuire}, {McIver}, {McManus}, {McRae}, {McWilliams}, {Meacher},
  {Meadors}, {Mehmet}, {Mehta}, {Meidam}, {Melatos}, {Mendell}, {Mercer},
  {Mereni}, {Merilh}, {Merzougui}, {Meshkov}, {Messenger}, {Messick},
  {Metzdorff}, {Meyers}, {Miao}, {Michel}, {Michimura}, {Middleton},
  {Mikhailov}, {Milano}, {Miller}, {Miller}, {Millhouse}, {Mills},
  {Milovich-Goff}, {Minazzoli}, {Minenkov}, {Mio}, {Mishkin}, {Mishra},
  {Mistry}, {Mitra}, {Mitrofanov}, {Mitselmakher}, {Mittleman}, {Miyakawa},
  {Miyamoto}, {Miyazaki}, {Miyo}, {Miyoki}, {Mo}, {Moffa}, {Mogushi},
  {Mohapatra}, {Montani}, {Moore}, {Moraru}, {Moreno}, {Morisaki}, {Moriwaki},
  {Mours}, {Mow-Lowry}, {Mukherjee}, {Mukherjee}, {Mukherjee}, {Mukund},
  {Mullavey}, {Munch}, {Mu{\~n}iz}, {Muratore}, {Murray}, {Nagano}, {Nagano},
  {Nagar}, {Nakamura}, {Nakano}, {Nakano}, {Nakashima}, {Nardecchia},
  {Narikawa}, {Naticchioni}, {Nayak}, {Negishi}, {Neilson}, {Nelemans},
  {Nelson}, {Nery}, {Neunzert}, {Ng}, {Ng}, {Nguyen}, {Ni}, {Nichols},
  {Nishizawa}, {Nissanke}, {Nocera}, {North}, {Nuttall}, {Obergaulinger},
  {Oberling}, {O'Brien}, {Obuchi}, {O'Dea}, {Ogaki}, {Ogin}, {Oh}, {Oh},
  {Ohashi}, {Ohishi}, {Ohkawa}, {Ohme}, {Ohta}, {Okada}, {Okutomi}, {Oliver},
  {Oohara}, {Ooi}, {Oppermann}, {Oram}, {O'Reilly}, {Ormiston}, {Ortega},
  {O'Shaughnessy}, {Oshino}, {Ossokine}, {Ottaway}, {Overmier}, {Owen}, {Pace},
  {Pagano}, {Page}, {Pai}, {Pai}, {Palamos}, {Palashov}, {Palomba},
  {Pal-Singh}, {Pan}, {Pan}, {Pang}, {Pang}, {Pang}, {Pankow}, {Pannarale},
  {Pant}, {Paoletti}, {Paoli}, {Papa}, {Parida}, {Park}, {Parker}, {Pascucci},
  {Pasqualetti}, {Passaquieti}, {Passuello}, {Patil}, {Patricelli},
  {Pearlstone}, {Pedersen}, {Pedraza}, {Pedurand}, {Pele}, {Arellano}, {Penn},
  {Perez}, {Perreca}, {Pfeiffer}, {Phelps}, {Phukon}, {Piccinni}, {Pichot},
  {Piergiovanni}, {Pillant}, {Pinard}, {Pinto}, {Pirello}, {Pitkin},
  {Poggiani}, {Pong}, {Ponrathnam}, {Popolizio}, {Porter}, {Powell},
  {Prajapati}, {Prasad}, {Prasai}, {Prasanna}, {Pratten}, {Prestegard},
  {Privitera}, {Prodi}, {Prokhorov}, {Puncken}, {Punturo}, {Puppo},
  {P{\"u}rrer}, {Qi}, {Quetschke}, {Quinonez}, {Quintero}, {Quitzow-James},
  {Raab}, {Radkins}, {Radulescu}, {Raffai}, {Raja}, {Rajan}, {Rajbhandari},
  {Rakhmanov}, {Ramirez}, {Ramos-Buades}, {Rana}, {Rao}, {Rapagnani},
  {Raymond}, {Razzano}, {Read}, {Regimbau}, {Rei}, {Reid}, {Reitze}, {Ren},
  {Ricci}, {Richardson}, {Richardson}, {Ricker}, {Riles}, {Rizzo}, {Robertson},
  {Robie}, {Robinet}, {Rocchi}, {Rolland}, {Rollins}, {Roma}, {Romanelli},
  {Romano}, {Romel}, {Romie}, {Rose}, {Rosi{\'n}ska}, {Rosofsky}, {Ross},
  {Rowan}, {R{\"u}diger}, {Ruggi}, {Rutins}, {Ryan}, {Sachdev}, {Sadecki},
  {Sago}, {Saito}, {Saito}, {Sakai}, {Sakai}, {Sakamoto}, {Sakellariadou},
  {Sakuno}, {Salconi}, {Saleem}, {Samajdar}, {Sammut}, {Sanchez}, {Sanchez},
  {Sanchis-Gual}, {Sandberg}, {Sanders}, {Santiago}, {Sarin}, {Sassolas},
  {Sathyaprakash}, {Sato}, {Sato}, {Sauter}, {Savage}, {Sawada}, {Schale},
  {Scheel}, {Scheuer}, {Schmidt}, {Schnabel}, {Schofield}, {Sch{\"o}nbeck},
  {Schreiber}, {Schulte}, {Schutz}, {Schwalbe}, {Scott}, {Scott}, {Seidel},
  {Sekiguchi}, {Sekiguchi}, {Sellers}, {Sengupta}, {Sennett}, {Sentenac},
  {Sequino}, {Sergeev}, {Setyawati}, {Shaddock}, {Shaffer}, {Shahriar},
  {Shaner}, {Shao}, {Sharma}, {Shawhan}, {Shen}, {Shibagaki}, {Shimizu},
  {Shimoda}, {Shimode}, {Shink}, {Shinkai}, {Shishido}, {Shoda}, {Shoemaker},
  {Shoemaker}, {Shyamsundar}, {Siellez}, {Sieniawska}, {Sigg}, {Silva},
  {Singer}, {Singh}, {Singhal}, {Sintes}, {Sitmukhambetov}, {Skliris},
  {Slagmolen}, {Slaven-Blair}, {Smith}, {Smith}, {Somala}, {Somiya}, {Son},
  {Sorazu}, {Sorrentino}, {Sotani}, {Souradeep}, {Sowell}, {Spencer},
  {Srivastava}, {Srivastava}, {Staats}, {Stachie}, {Standke}, {Steer},
  {Steinke}, {Steinlechner}, {Steinlechner}, {Steinmeyer}, {Stevenson},
  {Stocks}, {Stone}, {Stops}, {Strain}, {Stratta}, {Strigin}, {Strunk},
  {Sturani}, {Stuver}, {Sudhir}, {Sugimoto}, {Summerscales}, {Sun}, {Sunil},
  {Suresh}, {Sutton}, {Suzuki}, {Suzuki}, {Swinkels}, {Szczepa{\'n}czyk},
  {Tacca}, {Tagoshi}, {Tait}, {Takahashi}, {Takahashi}, {Takamori}, {Takano},
  {Takeda}, {Takeda}, {Talbot}, {Talukder}, {Tanaka}, {Tanaka}, {Tanaka},
  {Tanaka}, {Tanaka}, {Tanioka}, {Tanner}, {T{\'a}pai}, {Tapia San Martin},
  {Taracchini}, {Tasson}, {Taylor}, {Telada}, {Thies}, {Thomas}, {Thomas},
  {Thondapu}, {Thorne}, {Thrane}, {Tiwari}, {Tiwari}, {Tiwari}, {Toland},
  {Tomaru}, {Tomigami}, {Tomura}, {Tonelli}, {Tornasi}, {Torres-Forn{\'e}},
  {Torrie}, {T{\"o}yr{\"a}}, {Travasso}, {Traylor}, {Tringali}, {Trovato},
  {Trozzo}, {Trudeau}, {Tsang}, {Tsang}, {Tse}, {Tso}, {Tsubono}, {Tsuchida},
  {Tsukada}, {Tsuna}, {Tsuzuki}, {Tuyenbayev}, {Uchikata}, {Uchiyama}, {Ueda},
  {Uehara}, {Ueno}, {Ueshima}, {Ugolini}, {Unnikrishnan}, {Uraguchi}, {Urban},
  {Ushiba}, {Usman}, {Vahlbruch}, {Vajente}, {Valdes}, {van Bakel}, {van
  Beuzekom}, {van den Brand}, {van den Broeck}, {Vander-Hyde}, {van der
  Schaaf}, {van Heijningen}, {van Putten}, {van Veggel}, {Vardaro}, {Varma},
  {Vass}, {Vas{\'u}th}, {Vecchio}, {Vedovato}, {Veitch}, {Veitch},
  {Venkateswara}, {Venugopalan}, {Verkindt}, {Vetrano}, {Vicer{\'e}}, {Viets},
  {Vine}, {Vinet}, {Vitale}, {Vivanco}, {Vo}, {Vocca}, {Vorvick}, {Vyatchanin},
  {Wade}, {Wade}, {Wade}, {Walet}, {Walker}, {Wallace}, {Walsh}, {Wang},
  {Wang}, {Wang}, {Wang}, {Wang}, {Wang}, {Ward}, {Warden}, {Warner}, {Was},
  {Watchi}, {Weaver}, {Wei}, {Weinert}, {Weinstein}, {Weiss}, {Wellmann},
  {Wen}, {Wessel}, {We{\ss}els}, {Westhouse}, {Wette}, {Whelan}, {Whiting},
  {Whittle}, {Wilken}, {Williams}, {Williamson}, {Willis}, {Willke}, {Wimmer},
  {Winkler}, {Wipf}, {Wittel}, {Woan}, {Woehler}, {Wofford}, {Worden},
  {Wright}, {Wu}, {Wu}, {Wu}, {Wu}, {Wysocki}, {Xiao}, {Xu}, {Yamada},
  {Yamamoto}, {Yamamoto}, {Yamamoto}, {Yamamoto}, {Yancey}, {Yang}, {Yap},
  {Yazback}, {Yeeles}, {Yokogawa}, {Yokoyama}, {Yokozawa}, {Yoshioka}, {Yu},
  {Yu}, {Yuen}, {Yuzurihara}, {Yvert}, {Zadro{\.z}ny}, {Zanolin}, {Zeidler},
  {Zelenova}, {Zendri}, {Zevin}, {Zhang}, {Zhang}, {Zhang}, {Zhao}, {Zhao},
  {Zhou}, {Zhou}, {Zhu}, {Zhu}, {Zimmerman}, {Zucker}, {Zweizig}, {Kagra
  Collaboration}, \& {VIRGO Collaboration}}]{2020LRR....23....3A}
---. 2020, Living Reviews in Relativity, 23, 3,
  \dodoi{10.1007/s41114-020-00026-9}

\bibitem[{{An} {et~al.}(2021){An}, {Sun}, {Zhang}, {Yang}, {Li}, {Wen}, {Gong},
  {Liang}, {Liu}, {Liu}, {Li}, {Xiong}, {Xu}, {Zhang}, {Zhao}, {Cai}, {Chang},
  {Chen}, {Chen}, {Du}, {Feng}, {Gao}, {Gao}, {Guo}, {He}, {Hou}, {Li}, {Li},
  {Li}, {Li}, {Li}, {Lu}, {Lu}, {Meng}, {Peng}, {Shi}, {Wang}, {Wang}, {Wang},
  {Wang}, {Wen}, {Xiao}, {Xu}, {Yang}, {Yi}, {Zhang}, {Zhang}, {Zhang},
  {Zhang}, {Zhao}, \& {Zhou}}]{2021arXiv211204774A}
{An}, Z.~H., {Sun}, X.~L., {Zhang}, D.~L., {et~al.} 2021, arXiv e-prints,
  arXiv:2112.04774.
\newblock \doarXiv{2112.04774}

\bibitem[{{Aptekar} {et~al.}(1995){Aptekar}, {Frederiks}, {Golenetskii},
  {Ilynskii}, {Mazets}, {Panov}, {Sokolova}, {Terekhov}, {Sheshin}, {Cline}, \&
  {Stilwell}}]{1995SSRv...71..265A}
{Aptekar}, R.~L., {Frederiks}, D.~D., {Golenetskii}, S.~V., {et~al.} 1995,
  \ssr, 71, 265, \dodoi{10.1007/BF00751332}

\bibitem[{{Barthelmy} {et~al.}(2005){Barthelmy}, {Barbier}, {Cummings},
  {Fenimore}, {Gehrels}, {Hullinger}, {Krimm}, {Markwardt}, {Palmer},
  {Parsons}, {Sato}, {Suzuki}, {Takahashi}, {Tashiro}, \&
  {Tueller}}]{2005SSRv..120..143B}
{Barthelmy}, S.~D., {Barbier}, L.~M., {Cummings}, J.~R., {et~al.} 2005, \ssr,
  120, 143, \dodoi{10.1007/s11214-005-5096-3}

\bibitem[{{Biltzinger} {et~al.}(2021{\natexlab{a}}){Biltzinger}, {Kunzweiler},
  {Berlato}, {Greiner}, \& {Burgess}}]{2021GCN.30333....1B}
{Biltzinger}, B., {Kunzweiler}, F., {Berlato}, F., {Greiner}, J., \& {Burgess},
  J. 2021{\natexlab{a}}, GRB Coordinates Network, 30333, 1

\bibitem[{{Biltzinger} {et~al.}(2021{\natexlab{b}}){Biltzinger}, {Kunzweiler},
  {Berlato}, {Greiner}, \& {Burgess}}]{2021GCN.31033....1B}
---. 2021{\natexlab{b}}, GRB Coordinates Network, 31033, 1

\bibitem[{{Biltzinger} {et~al.}(2021{\natexlab{c}}){Biltzinger}, {Kunzweiler},
  {Berlato}, {Greiner}, \& {Burgess}}]{2021GCN.31239....1B}
---. 2021{\natexlab{c}}, GRB Coordinates Network, 31239, 1

\bibitem[{{Burgess} {et~al.}(2021){Burgess}, {Kunzweiler}, {Biltzinger},
  {Berlato}, \& {Greiner}}]{2021GCN.30727....1B}
{Burgess}, J., {Kunzweiler}, F., {Biltzinger}, B., {Berlato}, F., \& {Greiner},
  J. 2021, GRB Coordinates Network, 30727, 1

\bibitem[{{Burgess} {et~al.}(2018){Burgess}, {Yu}, {Greiner}, \&
  {Mortlock}}]{2018MNRAS.476.1427B}
{Burgess}, J.~M., {Yu}, H.-F., {Greiner}, J., \& {Mortlock}, D.~J. 2018,
  \mnras, 476, 1427, \dodoi{10.1093/mnras/stx2853}

\bibitem[{{Cai} {et~al.}(2021){Cai}, {Xiong}, {Li}, {Liu}, {Zhang}, {Li},
  {Song}, {Li}, {Xiao}, {Yi}, {Zhu}, {Zheng}, {Chen}, {Luo}, {Huang}, {Song},
  {Zhao}, {Zhao}, {Zhang}, {Bu}, {Cao}, {Chang}, {Chen}, {Chen}, {Chen},
  {Chen}, {Chen}, {Cui}, {Du}, {Gao}, {Gao}, {Ge}, {Gu}, {Guan}, {Guo}, {Han},
  {Huo}, {Jia}, {Jiang}, {Jin}, {Kong}, {Li}, {Li}, {Li}, {Li}, {Li}, {Li},
  {Liang}, {Liao}, {Liu}, {Liu}, {Liu}, {Liu}, {Lu}, {Lu}, {Luo}, {Ma}, {Ma},
  {Meng}, {Nang}, {Nie}, {Ou}, {Qu}, {Ren}, {Sai}, {Sun}, {Tan}, {Tao}, {Tuo},
  {Wang}, {Wang}, {Wang}, {Wang}, {Wang}, {Wen}, {Wu}, {Wu}, {Wu}, {Xiao},
  {Xu}, {Yang}, {Yang}, {Yang}, {Yang}, {Yang}, {Yin}, {You}, {Zhang}, {Zhang},
  {Zhang}, {Zhang}, {Zhang}, {Zhang}, {Zhang}, {Zhang}, {Zhang}, {Zhao},
  {Zheng}, \& {Zhou}}]{2021MNRAS.508.3910C}
{Cai}, C., {Xiong}, S.~L., {Li}, C.~K., {et~al.} 2021, \mnras, 508, 3910,
  \dodoi{10.1093/mnras/stab2760}

\bibitem[{{Chen} {et~al.}(2020){Chen}, {Li}, {Huang}, {Duan}, {Zhang}, {Guo},
  {Sun}, {Wang}, {Song}, {Li}, {Li}, {Ou}, {Zhao}, {Peng}, {Shi}, {Li}, {Li},
  {Xiao}, {Song}, {Wang}, {Ma}, {Zhang}, {Xiong}, {Cai}, {Zhang}, {Chen},
  {Qiao}, {Yao}, {Zheng}, \& {Zhao}}]{2020SSPMA..50l9512C}
{Chen}, W., {Li}, B., {Huang}, Y., {et~al.} 2020, Scientia Sinica Physica,
  Mechanica \& Astronomica, 50, 129512, \dodoi{10.1360/SSPMA-2020-0389}

\bibitem[{{Chen} {et~al.}(2022){Chen}, {Li}, {Xiong}, {Ji}, {Zhang}, {Peng},
  {Qiao}, {Li}, {Wen}, {Song}, {Zheng}, {Song}, {Zhao}, {Huang}, {Lu}, {Zhang},
  {Xiao}, {Cai}, {An}, {Chang}, {Chen}, {Chen}, {Chen}, {Dai}, {Du}, {Gao},
  {Gong}, {Guo}, {Guo}, {He}, {Li}, {Li}, {Li}, {Li}, {Li}, {Li}, {Li}, {Li},
  {Li}, {Liang}, {Liang}, {Liao}, {Liu}, {Liu}, {Liu}, {Luo}, {Ma}, {Meng},
  {Ou}, {Shi}, {Shi}, {Sun}, {Sun}, {Tuo}, {Wang}, {Wang}, {Wang}, {Wang},
  {Wang}, {Wang}, {Wang}, {Wang}, {Wen}, {Wu}, {Xie}, {Xu}, {Xu}, {Xue},
  {Yang}, {Yao}, {Ye}, {Yi}, {Zhang}, {Zhang}, {Zhang}, {Zhang}, {Zhang},
  {Zhang}, {Zhang}, {Zhang}, {Zhang}, {Zhang}, {Zhang}, {Zhao}, {Zhao}, {Zhao},
  {Zheng}, {Zhou}, \& {Zhu}}]{2021arXiv211204790C1}
{Chen}, Y.-P., {Li}, J., {Xiong}, S.-L., {et~al.} 2022, \apj, 935, 10,
  \dodoi{10.3847/1538-4357/ac7ff8}

\bibitem[{{D'Avanzo} {et~al.}(2021){D'Avanzo}, {Bernardini}, {Lien},
  {Melandri}, {Page}, {Palmer}, {Sbarrato}, \& {Neil Gehrels Swift Observatory
  Team}}]{2021GCN.30261....1D}
{D'Avanzo}, P., {Bernardini}, M.~G., {Lien}, A.~Y., {et~al.} 2021, GRB
  Coordinates Network, 30261, 1

\bibitem[{{DeLaunay} {et~al.}(2021){DeLaunay}, {Tohuvavohu}, {Raman}, \&
  {Kennea}}]{2021GCN.31047....1D}
{DeLaunay}, J., {Tohuvavohu}, A., {Raman}, G., \& {Kennea}, J.~A. 2021, GRB
  Coordinates Network, 31047, 1

\bibitem[{{Di Lalla} {et~al.}(2021){Di Lalla}, {Arimoto}, {Axelsson},
  {Kocevski}, {Omodei}, \& {Fermi-LAT Collaboration}}]{2021GCN.30062....1D}
{Di Lalla}, N., {Arimoto}, M., {Axelsson}, M., {et~al.} 2021, GRB Coordinates
  Network, 30062, 1

\bibitem[{{Fermi GBM Team}(2021{\natexlab{a}})}]{2021GCN.29659....1F}
{Fermi GBM Team}. 2021{\natexlab{a}}, GRB Coordinates Network, 29659, 1

\bibitem[{{Fermi GBM Team}(2021{\natexlab{b}})}]{2021GCN.29739....1F}
---. 2021{\natexlab{b}}, GRB Coordinates Network, 29739, 1

\bibitem[{{Fermi GBM Team}(2021{\natexlab{c}})}]{2021GCN.30298....1F}
---. 2021{\natexlab{c}}, GRB Coordinates Network, 30298, 1

\bibitem[{{Fermi GBM Team}(2021{\natexlab{d}})}]{2021GCN.30880....1F}
---. 2021{\natexlab{d}}, GRB Coordinates Network, 30880, 1

\bibitem[{{Fermi GBM Team}(2021{\natexlab{e}})}]{2021GCN.31207....1F}
---. 2021{\natexlab{e}}, GRB Coordinates Network, 31207, 1

\bibitem[{{Fermi GBM Team}(2021{\natexlab{f}})}]{2021GCN.31244....1F}
---. 2021{\natexlab{f}}, GRB Coordinates Network, 31244, 1

\bibitem[{{Fermi GBM Team}(2021{\natexlab{g}})}]{2021GCN.31336....1F}
---. 2021{\natexlab{g}}, GRB Coordinates Network, 31336, 1

\bibitem[{{Fermi GBM Team}(2021{\natexlab{h}})}]{2021GCN.31343....1F}
---. 2021{\natexlab{h}}, GRB Coordinates Network, 31343, 1

\bibitem[{{Goad} {et~al.}(2021){Goad}, {Osborne}, {Beardmore}, {Evans}, \&
  {Swift-XRT Team.}}]{2021GCN.30850....1G}
{Goad}, M.~R., {Osborne}, J.~P., {Beardmore}, A.~P., {Evans}, P.~A., \&
  {Swift-XRT Team.} 2021, GRB Coordinates Network, 30850, 1

\bibitem[{{Goldstein} {et~al.}(2020){Goldstein}, {Fletcher}, {Veres}, {Briggs},
  {Cleveland}, {Gibby}, {Hui}, {Bissaldi}, {Burns}, {Hamburg}, {Kienlin},
  {Kocevski}, {Mailyan}, {Malacaria}, {Paciesas}, {Roberts}, \&
  {Wilson-Hodge}}]{2020ApJ...895...40G}
{Goldstein}, A., {Fletcher}, C., {Veres}, P., {et~al.} 2020, \apj, 895, 40,
  \dodoi{10.3847/1538-4357/ab8bdb}

\bibitem[{{Goodman}(1986)}]{1986ApJ...308L..47G}
{Goodman}, J. 1986, \apjl, 308, L47, \dodoi{10.1086/184741}

\bibitem[{{Huang} {et~al.}(2021){Huang}, {Zheng}, {Lu}, {Zhang}, {Song},
  {Peng}, {Xiong}, {Xiao}, {Cai}, {Zhao}, {Ma}, {Zhang}, {Zhang}, {An}, {Chen},
  {Chen}, {Chen}, {Gao}, {Gong}, {Guo}, {He}, {Li}, {Li}, {Li}, {Li}, {Li},
  {Li}, {Li}, {Li}, {Liang}, {Liao}, {Liu}, {Liu}, {Liu}, {Luo}, {Ou}, {Peng},
  {Qiao}, {Shi}, {Shi}, {Song}, {Sun}, {Sun}, {Tuo}, {Wang}, {Wang}, {Wang},
  {Wen}, {Xu}, {Xu}, {Xue}, {Yang}, {Yao}, {Yi}, {Zhang}, {Zhang}, {Zhang},
  {Zhang}, {Zhang}, {Zhang}, {Zhang}, {Zhang}, {Zhang}, {Zhao}, {Zhao},
  {Zheng}, {Zhou}, \& {Gecam Team}}]{2021GCN.29363....1H}
{Huang}, Y., {Zheng}, S.~J., {Lu}, F.~J., {et~al.} 2021, GRB Coordinates
  Network, 29363, 1

\bibitem[{{Hurley} {et~al.}(2021{\natexlab{a}}){Hurley}, {Ipn}, {Mitrofanov},
  {Golovin}, {Litvak}, {Sanin}, {Hend-Odyssey Grb Team}, {Kozlova},
  {Golenetskii}, {Aptekar}, {Frederiks}, {Svinkin}, {Cline}, {Konus-Wind Team},
  {von Kienlin}, {Zhang}, {Rau}, {Savchenko}, {Bozzo}, {Ferrigno}, {INTEGRAL
  SPI-ACS Grb Team}, {Boynton}, {Fellows}, {Harshman}, {Enos}, {Starr}, \&
  {Grs-Odyssey Grb Team}}]{2021GCN.29997....1H}
{Hurley}, K., {Ipn}, {Mitrofanov}, I.~G., {et~al.} 2021{\natexlab{a}}, GRB
  Coordinates Network, 29997, 1

\bibitem[{{Hurley} {et~al.}(2021{\natexlab{b}}){Hurley}, {Ipn}, {Mitrofanov},
  {Golovin}, {Litvak}, {Sanin}, {Hend-Odyssey Grb Team}, {Kozlova},
  {Golenetskii}, {Aptekar}, {Frederiks}, {Svinkin}, {Cline}, {Konus-Wind Team},
  {von Kienlin}, {Zhang}, {Rau}, {Savchenko}, {Bozzo}, {Ferrigno}, {INTEGRAL
  SPI-ACS Grb Team}, {Boynton}, {Fellows}, {Harshman}, {Enos}, {Starr}, \&
  {Grs-Odyssey Grb Team}}]{2021GCN.30002....1H}
---. 2021{\natexlab{b}}, GRB Coordinates Network, 30002, 1

\bibitem[{{Hurley} {et~al.}(2021{\natexlab{c}}){Hurley}, {Ipn}, {Mitrofanov},
  {Golovin}, {Litvak}, {Sanin}, {Hend-Odyssey Grb Team}, {Kozlova},
  {Golenetskii}, {Aptekar}, {Frederiks}, {Svinkin}, {Cline}, {Konus-Wind Team},
  {von Kienlin}, {Zhang}, {Rau}, {Savchenko}, {Bozzo}, {Ferrigno}, {INTEGRAL
  SPI-ACS Grb Team}, {Boynton}, {Fellows}, {Harshman}, {Enos}, {Starr}, \&
  {Grs-Odyssey Grb Team}}]{2021GCN.30154....1H}
---. 2021{\natexlab{c}}, GRB Coordinates Network, 30154, 1

\bibitem[{{Kennea} {et~al.}(2021){Kennea}, {Osborne}, {Page}, {Ambrosi},
  {Capalbi}, {Perri}, {Tohuvavohu}, {Sbarufatti}, {Burrows}, {Beardmore},
  {Evans}, \& {Swift-XRT Team}}]{2021GCN.29413....1K}
{Kennea}, J.~A., {Osborne}, J.~P., {Page}, K.~L., {et~al.} 2021, GRB
  Coordinates Network, 29413, 1

\bibitem[{{Kozyrev} {et~al.}(2021{\natexlab{a}}){Kozyrev}, {Golovin}, {Litvak},
  {Mitrofanov}, {Sanin}, {Mgns/Bepicolombo Team}, {Hend/Mars Odyssey Team},
  {Benkhoff}, {Bepicolombo Team}, {Hurley}, {Ipn}, {Svinkin}, {Golenetskii},
  {Frederiks}, {Ridnaia}, {Lysenko}, {Cline}, {Konus-Wind Team}, {Goldstein},
  {Briggs}, {Wilson-Hodge}, {Fermi Gbm Team}, {von Kienlin}, {Zhang}, {Rau},
  {Savchenko}, {Bozzo}, {Ferrigno}, {INTEGRAL SPI-ACS Grb Team}, {Barthelmy},
  {Cummings}, {Krimm}, {Palmer}, {Tohuvavohu}, {Swift-Bat Team}, {Boynton},
  {Fellows}, {Harshman}, {Enos}, {Starr}, \& {Grs-Odyssey Grb
  Team}}]{2021GCN.30956....1K}
{Kozyrev}, A.~S., {Golovin}, D.~V., {Litvak}, M.~L., {et~al.}
  2021{\natexlab{a}}, GRB Coordinates Network, 30956, 1

\bibitem[{{Kozyrev} {et~al.}(2021{\natexlab{b}}){Kozyrev}, {Golovin}, {Litvak},
  {Mitrofanov}, {Sanin}, {Mgns/Bepicolombo Team}, {Hend/Mars Odyssey Team},
  {Benkhoff}, {Bepicolombo Team}, {Hurley}, {Ipn}, {Svinkin}, {Golenetskii},
  {Frederiks}, {Ridnaia}, {Lysenko}, {Cline}, {Konus-Wind Team}, {von Kienlin},
  {Zhang}, {Rau}, {Savchenko}, {Bozzo}, {Ferrigno}, {INTEGRAL SPI-ACS Grb
  Team}, {Barthelmy}, {Cummings}, {Krimm}, {Palmer}, {Tohuvavohu}, {Swift-Bat
  Team}, {Boynton}, {Fellows}, {Harshman}, {Enos}, \&
  {Starr}}]{2021GCN.31024....1K}
---. 2021{\natexlab{b}}, GRB Coordinates Network, 31024, 1

\bibitem[{{Kozyrev} {et~al.}(2021{\natexlab{c}}){Kozyrev}, {Golovin}, {Litvak},
  {Mitrofanov}, {Sanin}, {MGNS/BepiColombo}, {HEND/Mars Odyssey Teams},
  {Benkhoff}, {Hurley}, {IPN}, {Svinkin}, {Golenetskii}, {Frederiks},
  {Ridnaia}, {Lysenko}, {Cline}, {Konus-Wind Team}, {Boynton}, {Fellows},
  {Harshman}, {Enos}, {Starr}, \& {GRS-Okyssey GRB Team}}]{2021GCN.31129....1K}
---. 2021{\natexlab{c}}, GRB Coordinates Network, 31129, 1

\bibitem[{{Kozyrev} {et~al.}(2021{\natexlab{d}}){Kozyrev}, {Golovin}, {Litvak},
  {Mitrofanov}, {Sanin}, {Mgns/Bepicolombo Team}, {Hend/Mars Odyssey Team},
  {Benkhoff}, {Bepicolombo Team}, {Hurley}, {Ipn}, {Svinkin}, {Golenetskii},
  {Frederiks}, {Ridnaia}, {Lysenko}, {Cline}, {Konus-Wind Team}, {von Kienlin},
  {Zhang}, {Rau}, {Savchenko}, {Bozzo}, {Ferrigno}, {INTEGRAL SPI-ACS Grb
  Team}, {Barthelmy}, {Cummings}, {Krimm}, {Palmer}, {Tohuvavohu}, {Swift-Bat
  Team}, {Boynton}, {Fellows}, {Harshman}, {Enos}, \&
  {Starr}}]{2021GCN.31177....1K}
---. 2021{\natexlab{d}}, GRB Coordinates Network, 31177, 1

\bibitem[{{Kunzweiler} {et~al.}(2021){Kunzweiler}, {Biltzinger}, {Berlato},
  {Greiner}, \& {J.}}]{2021GCN.29354....1K}
{Kunzweiler}, F., {Biltzinger}, B., {Berlato}, F., {Greiner}, J.~B., \& {J.}
  2021, GRB Coordinates Network, 29354, 1

\bibitem[{{Laros} {et~al.}(1997){Laros}, {Boynton}, {Hurley}, {Kouveliotou},
  {McCollough}, {Fishman}, {Meegan}, {Palmer}, {Cline}, {Starr}, {Trombka},
  {J.~I.}, {Boer}, {Niel}, \& {Metzger}}]{1997ApJS..110..157L}
{Laros}, J.~G., {Boynton}, W.~V., {Hurley}, K.~C., {et~al.} 1997, \apjs, 110,
  157, \dodoi{10.1086/312993}

\bibitem[{{Li} {et~al.}(2021{\natexlab{a}}){Li}, {Guo}, {Lv}, {Zhao}, \&
  {He}}]{2021AdSpR..67.1701L}
{Li}, G., {Guo}, S., {Lv}, J., {Zhao}, K., \& {He}, Z. 2021{\natexlab{a}},
  Advances in Space Research, 67, 1701, \dodoi{10.1016/j.asr.2020.12.011}

\bibitem[{{Li} {et~al.}(1999){Li}, {Feng}, \& {Chen}}]{1999ApJ...521..789L}
{Li}, T.~P., {Feng}, Y.~X., \& {Chen}, L. 1999, \apj, 521, 789,
  \dodoi{10.1086/307566}

\bibitem[{{Li} {et~al.}(2021{\natexlab{b}}){Li}, {Wen}, {Xiong}, {Gong},
  {Zhang}, {An}, {Xu}, {Liu}, {Cai}, {Chang}, {Chen}, {Chen}, {Du}, {Gao},
  {Gao}, {Guo}, {He}, {Hou}, {Li}, {Li}, {Li}, {Li}, {Li}, {Li}, {Li}, {Li},
  {Liang}, {Liu}, {Lu}, {Lu}, {Ma}, {Meng}, {Peng}, {Qiao}, {Shi}, {Song},
  {Sun}, {Wang}, {Wang}, {Wang}, {Wang}, {Wang}, {Wen}, {Xiao}, {Xu}, {Yang},
  {Yang}, {Yi}, {Zhang}, {Zhang}, {Zhang}, {Zhang}, {Zhang}, {Zhang}, {Zhang},
  {Zhao}, {Zhao}, {Zheng}, \& {Zhou}}]{2021arXiv211204772L}
{Li}, X.~Q., {Wen}, X.~Y., {Xiong}, S.~L., {et~al.} 2021{\natexlab{b}}, arXiv
  e-prints, arXiv:2112.04772.
\newblock \doarXiv{2112.04772}

\bibitem[{{Li} {et~al.}(2020){Li}, {Wen}, {Sun}, {Liu}, {Liang}, {Guo}, {Peng},
  {Gong}, {Li}, {Wang}, {Xiong}, {Liao}, {Lu}, {Wang}, {An}, {Zhang}, {Gao},
  {Chen}, {Liu}, {Yang}, {Qiao}, {Zhang}, {Zhao}, {Xu}, {Zhu}, \&
  {Li}}]{2020SSPMA..50l9508L}
{Li}, Y., {Wen}, X., {Sun}, X., {et~al.} 2020, Scientia Sinica Physica,
  Mechanica \& Astronomica, 50, 129508, \dodoi{10.1360/SSPMA-2019-0417}

\bibitem[{{Lipunov} {et~al.}(2021{\natexlab{a}}){Lipunov}, {Kornilov},
  {Gorbovskoy}, {Tiurina}, {Balanutsa}, {Balakin}, {Vladimirov}, {Kuznetsov},
  {Zhirkov}, {Vlasenko}, {Gorbunov}, {Zimnukhov}, {Senik}, {Chasovnikov},
  {Pozdnyakov}, {Topolev}, {Cheryasov}, {Podesta}, {Lopez}, {Podesta},
  {Francile}, {Rebolo}, {Serra}, {Buckley}, {Gres}, {Budnev}, {Tlatov},
  {Dormidontov}, {Gabovich}, {Yurkov}, \& {Sergienko}}]{2021GCN.29339....1L}
{Lipunov}, V., {Kornilov}, V., {Gorbovskoy}, E., {et~al.} 2021{\natexlab{a}},
  GRB Coordinates Network, 29339, 1

\bibitem[{{Lipunov} {et~al.}(2021{\natexlab{b}}){Lipunov}, {Gorbovskoy},
  {Zhirkov}, {Tiurina}, {Balanutsa}, {Kuznetsov}, {Gress}, {Antipov},
  {Vlasenko}, {Senik}, {Topolev}, {Chasovnikov}, {Minkina}, {Grinshpun},
  {Cheryasov}, {Rebolo}, {Serra}, {Buckley}, {Budnev}, {Francile}, {Podesta},
  {Podesta}, {Tlatov}, {Dormidontov}, {Gabovich}, \&
  {Yurkov}}]{2021GCN.31048....1L}
{Lipunov}, V., {Gorbovskoy}, E., {Zhirkov}, K., {et~al.} 2021{\natexlab{b}},
  GRB Coordinates Network, 31048, 1

\bibitem[{{Meegan} {et~al.}(2009){Meegan}, {Lichti}, {Bhat}, {Bissaldi},
  {Briggs}, {Connaughton}, {Diehl}, {Fishman}, {Greiner}, {Hoover}, {van der
  Horst}, {von Kienlin}, {Kippen}, {Kouveliotou}, {McBreen}, {Paciesas},
  {Preece}, {Steinle}, {Wallace}, {Wilson}, \&
  {Wilson-Hodge}}]{2009ApJ...702..791M}
{Meegan}, C., {Lichti}, G., {Bhat}, P.~N., {et~al.} 2009, \apj, 702, 791,
  \dodoi{10.1088/0004-637X/702/1/791}

\bibitem[{{Nakar}(2020)}]{2020PhR...886....1N}
{Nakar}, E. 2020, \physrep, 886, 1, \dodoi{10.1016/j.physrep.2020.08.008}

\bibitem[{{Paczynski}(1986)}]{1986ApJ...308L..43P}
{Paczynski}, B. 1986, \apjl, 308, L43, \dodoi{10.1086/184740}

\bibitem[{{Page} {et~al.}(2021){Page}, {Gropp}, {Kuin}, {Lien}, \& {Neil
  Gehrels Swift Observatory Team}}]{2021GCN.30677....1P}
{Page}, K.~L., {Gropp}, J.~D., {Kuin}, N.~P.~M., {Lien}, A.~Y., \& {Neil
  Gehrels Swift Observatory Team}. 2021, GRB Coordinates Network, 30677, 1

\bibitem[{{Tohuvavohu} {et~al.}(2021){Tohuvavohu}, {Raman}, {DeLaunay}, \&
  {Kennea}}]{2021GCN.31334....1T}
{Tohuvavohu}, A., {Raman}, G., {DeLaunay}, J., \& {Kennea}, J.~A. 2021, GRB
  Coordinates Network, 31334, 1

\bibitem[{{Winkler} {et~al.}(2003){Winkler}, {Courvoisier}, {Di Cocco},
  {Gehrels}, {Gim{\'e}nez}, {Grebenev}, {Hermsen}, {Mas-Hesse}, {Lebrun},
  {Lund}, {Palumbo}, {Paul}, {Roques}, {Schnopper}, {Sch{\"o}nfelder},
  {Sunyaev}, {Teegarden}, {Ubertini}, {Vedrenne}, \&
  {Dean}}]{2003A&A...411L...1W}
{Winkler}, C., {Courvoisier}, T.~J.~L., {Di Cocco}, G., {et~al.} 2003, \aap,
  411, L1, \dodoi{10.1051/0004-6361:20031288}

\bibitem[{{Xiao} {et~al.}(2021){Xiao}, {Xiong}, {Zhang}, {Song}, {Lu}, {Huang},
  {Cai}, {Yi}, {Song}, {Chen}, {Ge}, {Liu}, {Li}, {Li}, \&
  {Zhao}}]{2021ApJ...920...43X}
{Xiao}, S., {Xiong}, S.~L., {Zhang}, S.~N., {et~al.} 2021, \apj, 920, 43,
  \dodoi{10.3847/1538-4357/ac1420}

\bibitem[{{Xiao} {et~al.}(2022){Xiao}, {Xiong}, {Cai}, {Song}, {Zheng}, {Peng},
  {Wang}, {Qiao}, {Guo}, {Wang}, {Li}, {Song}, {Yuan}, {Fan}, {Zhao}, {Huang},
  {Ma}, {Zhang}, {Li}, {Ge}, {Tuo}, {Chen}, {Zhang}, {He}, {Li}, {Yi}, {Zhao},
  {Zhang}, {Zheng}, {Xue}, {Liu}, {Zhang}, {Li}, {Zhang}, {Zhao}, {Zhao},
  {Guo}, {Xie}, {Wang}, {Zhang}, {Wang}, {Li}, {Li}, {Zhang}, {Shi}, {Zhao},
  {Yao}, {An}, {Chen}, {Gong}, {Liu}, {Gao}, {Li}, {Li}, {Liang}, {Liu}, {Sun},
  {Wang}, {Wen}, {Xu}, {Xu}, {Yang}, {Zhang}, {Zhang}, {Zhang}, {Chen}, {Lu},
  {Sun}, {Zhang}, \& {Zhang}}]{2022MNRAS.tmp..994X}
{Xiao}, S., {Xiong}, S.-L., {Cai}, C., {et~al.} 2022, \mnras,
  \dodoi{10.1093/mnras/stac999}

\bibitem[{{Xu} {et~al.}(2021){Xu}, {Sun}, {Yang}, {Li}, {Peng}, {Gong},
  {Liang}, {Liu}, {Guo}, {Wang}, {Li}, {An}, {He}, {Liu}, {Xiong}, {Wen},
  {Zhang}, {Zhang}, {Zhao}, {Zhang}, {Cai}, {Chang}, {Chen}, {Chen}, {Du},
  {Gao}, {Gao}, {Hou}, {Li}, {Li}, {Li}, {Li}, {Li}, {Lu}, {Lu}, {Meng}, {Shi},
  {Wang}, {Wang}, {Wang}, {Wen}, {Xiao}, {Xu}, {Yang}, {Yi}, {Zhang}, {Zhang},
  {Zhang}, {Zhao}, \& {Zhou}}]{2021arXiv211205314X}
{Xu}, Y.~B., {Sun}, X.~L., {Yang}, S., {et~al.} 2021, arXiv e-prints,
  arXiv:2112.05314.
\newblock \doarXiv{2112.05314}

\bibitem[{Yang {et~al.}(2019)Yang, Gao, Guo, Mao, \&
  Yang}]{2019Navigation..66.7Y}
Yang, Y., Gao, W., Guo, S., Mao, Y., \& Yang, Y. 2019, NAVIGATION, 66, 7,
  \dodoi{https://doi.org/10.1002/navi.291}

\bibitem[{{Zhang} {et~al.}(2020){Zhang}, {Li}, {Lu}, {Song}, {Xu}, {Liu},
  {Chen}, {Cao}, {Bu}, {Chang}, {Chen}, {Chen}, {Chen}, {Chen}, {Chen}, {Cui},
  {Cui}, {Deng}, {Dong}, {Du}, {Fu}, {Gao}, {Gao}, {Gao}, {Ge}, {Gu}, {Guan},
  {Gungor}, {Guo}, {Han}, {Hu}, {Huang}, {Huo}, {Jia}, {Jiang}, {Jiang}, {Jin},
  {Jin}, {Li}, {Li}, {Li}, {Li}, {Li}, {Li}, {Li}, {Li}, {Li}, {Li}, {Li},
  {Liang}, {Liao}, {Liu}, {Liu}, {Liu}, {Liu}, {Liu}, {Liu}, {Lu}, {Lu}, {Luo},
  {Ma}, {Meng}, {Nang}, {Nie}, {Ou}, {Qu}, {Sai}, {Shang}, {Shen}, {Sun},
  {Tan}, {Tao}, {Tuo}, {Wang}, {Wang}, {Wang}, {Wang}, {Wang}, {Wang}, {Wang},
  {Wen}, {Wu}, {Wu}, {Wu}, {Xiao}, {Xiong}, {Yan}, {Yang}, {Yang}, {Yang},
  {Yi}, {Yuan}, {Zhang}, {Zhang}, {Zhang}, {Zhang}, {Zhang}, {Zhang}, {Zhang},
  {Zhang}, {Zhang}, {Zhang}, {Zhang}, {Zhang}, {Zhang}, {Zhang}, {Zhang},
  {Zhang}, {Zhang}, {Zhang}, {Zhang}, {Zhang}, {Zhao}, {Zhao}, {Zheng}, {Zhou},
  {Zhu}, {Zhu}, {Zhuang}, \& {Insight-HXMT Team}}]{2020SCPMA..6349502Z}
{Zhang}, S.-N., {Li}, T., {Lu}, F., {et~al.} 2020, Science China Physics,
  Mechanics, and Astronomy, 63, 249502, \dodoi{10.1007/s11433-019-1432-6}

\bibitem[{{Zhao} {et~al.}(2021){Zhao}, {Xiong}, {Wen}, {Li}, {Cai}, {Xiao},
  {Luo}, {Peng}, {Guo}, {An}, {Gong}, {Liao}, {Zhang}, {Huang}, {Li}, {Wen},
  {Zhang}, {Duan}, {Wang}, {Shi}, {Zhang}, {Yi}, {Li}, {Xu}, {Liang}, {Liu},
  {Zhang}, {Sun}, {Zhang}, {Chen}, {Wang}, {Yang}, {Liu}, {Gao}, {Li}, {Wang},
  {Zhou}, {Zhao}, {Xue}, {Zheng}, {Liu}, {Han}, {Qi}, {Huang}, {Zhang}, {Chen},
  {Yang}, {Hou}, {Wang}, {Qiao}, {Ma}, {Li}, {Wang}, {Song}, {Song}, {Zheng},
  {Li}, {Zhang}, {Zhu}, {Chen}, {He}, {Zhang}, {Hou}, {Wang}, {Hao}, {Wang},
  {Yang}, {Wen}, {Chang}, {Du}, {Gao}, {Lan}, {Li}, {Li}, {Li}, {Lu}, {Lu},
  {Meng}, {Shi}, {Wang}, {Wang}, {Xu}, {Yang}, {Yang}, {Zhang}, {Zhang},
  {Zhang}, {Tang}, \& {Cheng}}]{2021arXiv211205101Z}
{Zhao}, X.-Y., {Xiong}, S.-L., {Wen}, X.-Y., {et~al.} 2021, arXiv e-prints,
  arXiv:2112.05101.
\newblock \doarXiv{2112.05101}

\bibitem[{{Zheng} {et~al.}(submitted to RAA){Zheng}, {Song}, {Ma}, {Wang},
  {Qiao}, {Huang}, {Zhao}, {Zhang}, {Li}, \& {Xiong}}]{2021RAA_inprep_Z}
{Zheng}, S.-J., {Song}, L.-M., {Ma}, X., {et~al.} submitted to RAA

\end{thebibliography}
\bibliographystyle{aasjournal}

\end{document}